\documentclass[fleqn,usenatbib]{mnras}

\usepackage{amsmath}
\usepackage{amssymb}
\usepackage{color}
\usepackage{epic}
\usepackage{epstopdf}
\usepackage{graphicx}
\usepackage{hyperref}
\usepackage{multirow}
\usepackage{natbib}
\usepackage{overpic}
\usepackage{xcolor}
\usepackage{color}
\usepackage{listings}
\AtBeginShipout{%
 \ifnum\value{page}>1 %
 \typeout{* Additional boxing of page `\thepage'}%
 \setbox\AtBeginShipoutBox=\hbox{\copy\AtBeginShipoutBox}%
 \fi
}

\title[RAiSE\textcolor{purple}{Red}: radio continuum redshifts for lobed AGNs]{RAiSE\textcolor{purple}{Red}: radio continuum redshifts for lobed AGNs}

\author[R. J. Turner et al.]{
Ross J. Turner$^{1}$\thanks{Email: turner.rj@icloud.com},  Guillaume Drouart$^{2}$, Nick Seymour$^{2}$ and Stanislav S. Shabala$^{1,3}$\\
$^{1}$School of Natural Sciences, University of Tasmania, Private Bag 37, Hobart, 7001, Australia\\
$^{2}$International Centre for Radio Astronomy Research, Curtin University, Bentley, WA 6102, Australia\\
$^{3}$ARC Centre of Excellence for All-Sky Astrophysics in 3 Dimensions (ASTRO 3D)}

\date{Accepted 2020 October 1. Received 2020 September 9; in original form 2020 July 17.}

\pubyear{2020}

\begin{document}

\label{firstpage}
\pagerange{\pageref{firstpage}--\pageref{lastpage}}
\maketitle

\begin{abstract}

Next-generation radio surveys are expected to detect tens of millions of active galactic nuclei (AGN) with a median redshift of $z \geqslant 1$. Beyond targeted surveys, the vast majority of these objects will not have spectroscopic redshifts, whilst photometric redshifts for high-redshift AGNs are of limited quality, and even then require optical and infrared photometry. We propose a new approach to measure the redshifts of lobed radio galaxies based exclusively on radio-frequency imaging and broadband radio photometry. Specifically, our algorithm uses the lobe flux density, angular size and width, and spectral shape to derive probability density functions for the most likely source redshift based on the \emph{Radio AGN in Semi-analytic Environments} (RAiSE) dynamical model.
The full physically based model explains 70\% of the variation in the spectroscopic redshifts of a high-redshift ($2 < z < 4$) sample of radio AGNs, compared to at most 27\% for any one of the observed attributes in isolation. We find that upper bounds on the angular size, as expected for unresolved sources, are sufficient to yield accurate redshift measurements at $z \geqslant 2$. The error in the model upon calibration using at least nine sources with known spectroscopic redshifts is $<$\,14\% in redshift (as $1 + z$) across all redshifts. We provide \emph{python} code for the calculation and calibration of our radio continuum redshifts in an online library.

\end{abstract}


\section{INTRODUCTION}
\label{sec:INTRODUCTION}

Next-generation large-sky radio surveys (e.g. ASKAP EMU, \citealt{Norris+2011}; ASKAP POSSUM, \citealt{Gaensler+2010}; LOFAR LoTSS, \citealt{Shimwell+2017, Shimwell+2019}; MeerKAT MIGHTEE, \citealt{Jarvis+2012}; MWA GLEAM, \citealt{Wayth+2015}; VLA VLASS,  \citealt{Lacy+2020}) are expected to catalogue tens of millions of galaxies \citep{Norris+2017}. Redshift estimates for these sources will be crucial to achieve many of the science goals. However, spectroscopic data will not be available for the vast majority of sources in large-sky surveys whilst, in many cases, optical and infrared photometry is expected to be of limited quality \citep{Norris+2019}. Moreover, the active galactic nuclei (AGN) population tends to be located at higher redshifts (median $\geqslant 1$) than the `radio quiet' optical sources for which spectral energy distribution template fitting techniques are developed \citep[e.g.][]{Arnouts+1999, Duncan+2018a}. The necessary combination of host galaxy and AGN spectral templates (i.e. their type and relative strength) further complicates template-based photometric redshift techniques \citep[e.g.][]{Salvato+2018}. Photons from the accretion disk, inner jet and dust additionally modify the spectral energy distribution in high-excitation radio galaxies which comprise the majority of AGNs above redshift $z = 2$-3 \citep[e.g.][]{Fabian+2012}.
\citet{Norris+2013} argue that close to half the galaxy population observed by next-generation radio surveys will be AGNs, whilst up to 100 per cent of large galaxies at $z \approx 0$ are thought to host an active nucleus (with some level of activity) at their centre \citep{Sabater+2019}. This highlights the importance of finding an alternative measurement of redshift for these rather ubiquitous objects.

Supervised machine learning has been touted as a viable alternative to conventional template-fitting techniques \citep[see][for review]{Salvato+2018}. These methods use a `training set' of radio sources with known spectroscopic redshifts to find correlations between their attributes (e.g. observed photometry) and the redshift \citep{Firth+2002, Tagliaferri+2003}.
Machine learning based implementations include random forests, neural networks, Gaussian processes, and support vector machines, with the ability to generate probability density functions for the redshift \citep{Amaro+2018, Duncan+2018b}. \citet{Norris+2019} compare the performance of several machine learning algorithms to a conventional template-fitting method, in particular considering the effect of limited multi-wavelength data to supplement radio observations. The quality of photometric redshifts is found to degrade with the reducing quality in the optical, infrared, radio and X-ray observations, however some measure of the redshift is still possible in most sources. Importantly, machine learning techniques are found to perform significantly worse at higher redshifts ($z \gtrsim 1$) due to the lack of training data for high redshift radio sources. Further, this approach does not provide a solution for high-redshift extended radio sources whose host galaxies are often too faint to be detected.

The quasar and extended radio galaxy subclasses of AGN have attracted more physically-based approaches to construct standard candles/rulers, primarily for use in cosmology, since their discovery over five decades ago. 
\citet{Watson+2011} showed the distance, and thus redshift, of $z < 0.3$ quasars can be constrained using a known relationship between the optical luminosity and the size of the broad emission line region \citep[see also][]{Haas+2011, Czerny+2013, King+2014}. 
Optical quasars have also been standardised using a correlation between their luminosity and the time lag between the optical and dust continuum \citep[e.g.][]{Oknyanskij+1999, Honig+2014, Yoshii+2014}. Meanwhile, existing techniques to standardise extended radio galaxies generally rely on large sample statistics to find a meaningful relationship with spectroscopic distance measurements \citep[e.g.][]{Kellermann+1993, Daly+1994, Buchalter+1998, Jackson+2004}. %For example, the \emph{Extended Radio Galaxy} method of \citet{Daly+1994} assumes the jet-inflated lobes of radio sources have a physical limit on their size that is redshift invariant (this assumption is invalid due to selection biases and the cosmological evoultion of cluster environments). 
However, the maximum lobe size at a given redshift can only be found by sampling a population, and thus these techniques cannot be applied to estimate redshifts for individual objects.

\citet{Turner+2019} proposed combining radio imaging and broadband radio-frequency observations to successfully standardise extended AGNs from the present epoch to ultra-high redshifts ($0 < z < 7$). They modified the analytic theory underpinning the \emph{Radio AGNs in Semi-analytic Environments} \citep[RAiSE;][]{Turner+2015, Turner+2018a} model for the dynamical and synchrotron evolution of radio AGNs to physically link the spatial and spectral observations.  In that work, the Hubble constant was accurately constrained using measurements of the integrated flux density, angular size and width, and spectral shape for the two lobes of Cygnus A. Meanwhile, previous iterations of RAiSE have found success in: (1) reproducing surface brightness and spectral age maps for canonical FR-I (3C31) and FR-II (3C436) type sources \citep{Turner+2018a}; (2) deriving jet kinetic powers consistent with X-ray inverse-Compton measurements \citep{Turner+2018b}; and (3) finding dynamical evolution in the lobe length, axis ratio and volume consistent with hydrodynamical simulations \citep{Turner+2018b}. 
\citet{Turner+2019} suggested several other applications for their standardised extended AGNs, including: relative (i.e. uncalibrated) distance measurements of high-redshift sources to constrain the matter and dark energy densities; and radio continuum redshifts making use of a sample of objects with known spectroscopic redshifts to calibrate the model.

In this work, we extend the theory developed by \citet{Turner+2019} to constrain the redshifts of extended radio AGNs inhabiting cosmological environments described by a Bayesian prior probability density function (Section \ref{sec:RADIO CONTINUUM REDSHIFTS}); this \emph{RAiSE\textcolor{purple}{Red}} algorithm only requires multi-frequency radio observations to estimate radio continuum redshifts. 
In Section \ref{sec:ASSESSMENT OF METHOD ON MOCK SOURCES}, the sensitivity of the method to either large uncertainties or limits in the observed radio source attributes is assessed using a population of mock Cygnus A-like sources. The \emph{RAiSE\textcolor{purple}{Red}} algorithm is applied to a sample of 17 objects across a range of redshifts ($0 < z < 4$) in Section \ref{sec:APPLICATION TO HIGH-REDSHIFT SOURCES}; we assess both the error in the uncalibrated model and the error following calibration using moderate samples of sources with known spectroscopic redshifts. Finally, we make our concluding remarks about the potential applications of this work in Section \ref{sec:CONCLUSIONS}. We provide \emph{python} code for the calculation and calibration of our radio continuum redshifts in the online supplementary material.

The $\Lambda \rm CDM$ concordance cosmology with $\Omega_{\rm M} = 0.3089\pm0.0062$, $\Omega_\Lambda = 0.6911\pm0.0062$ and $H_0 = 67.74\pm0.46 \rm\,km \,s^{-1} \,Mpc^{-1}$ \citep{Planck+2016} is assumed throughout the paper. The spectral index $\alpha$ is defined in the form $S = \nu^{-\alpha}$ for flux density $S$ and frequency $\nu$.

\section{RADIO CONTINUUM REDSHIFTS}
\label{sec:RADIO CONTINUUM REDSHIFTS}

The powerful \citeauthor{FR+1974} type-II (FR-II) radio lobe morphology is well modelled both analytically \citep[e.g.][]{KA+1997, Blundell+1999, Turner+2015, Hardcastle+2018} and numerically \citep[e.g.][]{Krause+2012, Hardcastle+2014, Yates+2018, Massaglia+2019}. Radio-frequency observables including the flux density, angular size and shape of the spectral energy distribution have been used to constrain not only intrinsic source properties such as jet kinetic power and age \citep[e.g.][]{Turner+2018b}, but also the line-of-sight transverse comoving distance \citep{Turner+2019}.
The radio continuum redshift, $z^*$ (in the context of this work), is defined as the trial redshift, $z$, that yields dynamical model predicted transverse comoving distances, $d_{\rm M}(z)$, in closest agreement with the expectation for the concordance cosmological model, $d_{{\rm M}\:\! (H_0, \Omega_{\rm m})}(z)$. The dynamical model  can be further calibrated using radio AGNs with known spectroscopic redshifts to more confidently constrain the distance to similar sources lacking redshifts.

\subsection{Theoretical background}

The dynamical model-based estimate of the transverse comoving distance to an active, lobed radio source at a trial redshift $z$ can be expressed in terms of the flux density, $S_\nu$ (single lobe at observer-frame frequency $\nu$); angular size, $\theta$ (single lobe); axis ratio, $A$ (single lobe length divided by the lobe radius); and properties of the electron energy distribution. Particles injected into the lobe are initially described by a power law distribution of energies $N(E) = N_0 E^{-s}$, where $N_0$ is a constant and $s = 2\alpha_{\rm inj} + 1$ for injection-time spectral index $\alpha_{\rm inj} > 0.5$. The spectral index of the electron energy population steepens to $\alpha_{\rm inj} + 0.5$ above the `optically-thin' break frequency, $\nu_{\rm b}$, due to synchrotron radiative losses as the source ages. These parameters describing the electron population are constrained observationally by fitting the radio spectrum with the continuous injection (CI) model \citep{Turner+2018b}. However, the spectral shape of remnants and jetted FR-Is are not fitted well by the functional form of the CI model \citep[e.g. 3C28 and 3C31 in][]{Harwood+2017}, enabling these objects to be excluded even without high-resolution imaging. Meanwhile, the flux density, source size and axis ratio are readily measured from high-resolution radio images. Importantly, the monochromatic flux density, $S_\nu$, used in analytic models without a full treatment of radiative loss mechanisms \citep[cf. RAiSE;][]{Turner+2015} must be measured at a frequency below the break frequency ($\nu < \nu_{\rm b}$).

Following \citet{Turner+2019}, the expected distance to a lobed FR-II radio source at some trial redshift $z$ is given by:
\begin{equation}
\begin{split}
d_{\rm M}&(..., z) = \bigg[\frac{A^2 [10^{b_1} \gamma]^{2 - s}\nu^{(s - 1)/2} S_\nu}{f_1(s)}\bigg]^{-1/(2 - y)} \\
&\quad\times\, \left[[10^{b_2}  \rho a^\beta] f_3(A, B, \beta, z, b_3) \frac{q \nu_{\rm b}}{q + 1}\right]^{x/(2 - y)} \\
&\quad\times\, \theta^{y/(2 - y)} [1 + z]^{[x(1 + b_4) - y - 2]/(2 - y)} ,
\label{dM}
\end{split}
\end{equation}
where $q = u_{\rm B}/u_{\rm e} \sim 0.019$ is the ratio of the energy in the magnetic field to that in the particles \citep[i.e. equipartition factor;][]{Turner+2018b}, and $\gamma \sim 300$ is the minimum Lorentz factor of the electron population after expanding from the jet terminal hotspot to the lobe \citep{Turner+2019}; the maximum Lorentz factor is neglected in the model as it has no significant effect on our results for $\gamma_{\rm max} \gg \gamma$. The density profile, of the total gas mass, into which the radio source expands (see Section \ref{sec:Cosmological environments}) is approximated locally as a power law of the form $\rho(r) = \rho [r/a]^{-\beta}$; the gas density $\rho$ is defined at some galactocentric radius $a$, which we arbitrarily set as the physical equivalent of the angular size of the lobe at the trial redshift $z$. Meanwhile, the exponents $x$ and $y$ are functions of $s$, $\beta$ and the lobe magnetic field strength, $B$, defined immediately after Equation 13 in \citet{Turner+2019}. The lobe magnetic field strength at the trial redshift $z$ is estimated using the following equation derived from Equations 3, 12 and 13 of \citet{Alexander+2000}:
\begin{equation}
\begin{split}
B&(..., z) = \sqrt{2\mu_0\,} \bigg[\frac{A^2 [10^{b_1} \gamma]^{2 - s}\nu^{(s - 1)/2} S_\nu [1 + z]^{5}}{f_1(s)\:\! \theta^3\:\! d_{{\rm M}\:\! (H_0, \Omega_{\rm m})}(z)} \bigg]^{2/(s + 5)} ,
\label{Bfield}
\end{split}
\end{equation}
where $\mu_0$ is the vacuum permeability, and $d_{{\rm M}\:\! (H_0, \Omega_0)}(z)$ is the transverse comoving distance at the trial redshift $z$ for the concordance cosmological model.

The constants of proportionality $f_1$ and $f_3$ in the distance equation (Equation \ref{dM}) are functions of $s$, and $A$, $B$, $\beta$ and $z$, respectively defined as:
\begin{subequations}
\begin{equation}
f_1(s) = \frac{\sigma_{\rm T} [s - 2]}{9 m_{\rm e} c}\left[\frac{e^2 \mu_0}{2 \pi^2 {m_{\rm e}}^2}\right]^{(s - 3)/4} \overline{\mathcal{Y}(t, \nu)} ,
\label{f1}
\end{equation}
\vspace{-10pt}
\begin{equation}
f_3(A, B, \beta, z, b_3) = \frac{18 \chi(A, \beta, b_3) \kappa^4(B, z)}{\upsilon^2 [\Gamma_{\rm x} + 1][5 - \beta]^2 [2\mu_0]^{\sigma(B, z)}[\Gamma_{\rm c} - 1]} ,
\label{f3}
\end{equation}
\end{subequations}
where $\sigma_{\rm T}$ is the electron scattering cross-section, $e$ and $m_{\rm e}$ are the electron charge and mass, $c$ is the speed of light, $\upsilon$ is a constant defined in Equation 5 of \citet{Turner+2018b}, and $\Gamma_{\rm c} = 4/3$ and $\Gamma_{\rm x} = 5/3$ are the adiabatic indices of the lobe plasma and external medium respectively. The time-average of the synchrotron radiative loss function, $\overline{\mathcal{Y}(t, \nu)}$, is contrained to be a constant value in the range 0.3-0.5 based on RAiSE simulations \citep{Turner+2019}. Meanwhile, the functions $\kappa(B, z)$ and $\sigma(B, z)$, defined in Equation 10 of \citet{Turner+2020}, are used to maintain an analytic solution across all lobe magnetic field strengths.
The ratio of the lobe to expansion surface pressures is found from the numerical simulations of \citet{Kaiser+1999} to be well modelled as \citep[Equation 7 of][]{Kaiser+2000}:
\begin{equation}
\begin{split}
\chi(A, \beta, b_3) = \frac{1}{2.14 - 0.52\beta}\left[A/2 \right]^{0.25(\beta + b_3) - 2.04} .
\label{chi}
\end{split}
\end{equation}
The key model parameters that can be constrained through observations, either for individual sources or as a population average, are summarised in Table~\ref{tab:summary}.

\begin{table*}
\begin{center}
\caption[]{Summary of key model parameters that can be constrained through observations. The parameters are grouped into three categories: (1) `radio continuum attributes', which must be measured for each individual source (each lobe separately or as an average); (2) `cosmological environments', which are simulated at each trial redshift; and (3) `fixed parameters',  which take a population average for all sources.}
\label{tab:summary}
\renewcommand{\arraystretch}{1.1}
\setlength{\tabcolsep}{8pt}
\begin{tabular}{cccc}
\hline\hline
\multicolumn{4}{c}{Radio continuum attributes} 
\\
\hline
flux density&$S_\nu$&Janskys& \multirow{4}{*}{\vspace{-1pt}radio imaging} \\
frequency (observer-frame)&$\nu$&Hertz& \\
angular size&$\theta$&arcsec& \\
axis ratio&$A$&--& \\
\cline{4-4}
injection index&$s$&--& \multirow{2}{*}{\vspace{-4pt}CI model spectrum} \\
break frequency&$\nu_{\rm b}$&Hertz& \\
\hline\hline
\multicolumn{4}{c}{Cosmological environments} 
\\
\hline
{density at lobe tip}&$\rho$& $\rm kg/m^{3}$& \multirow{2}{*}{\vspace{-1pt}SAGE simulations} \\
{density exponent} &$\beta$& -- \\
\hline\hline
\multicolumn{4}{c}{Fixed parameters} 
\\
\hline
equipartition factor&$q$& 0.019 & \citet{Turner+2018b}\\
minimum Lorentz factor&$\gamma$& 300 &\citet{Turner+2019}\\
\hline
\end{tabular}
\end{center}
\end{table*}

The dynamical model-based estimate for the transverse comoving distance can be calibrated using a modest-sized sample of radio AGNs with known spectroscopic redshifts. Calibration constants are included in the distance equation (Equation \ref{dM}) to provide small corrections to the most poorly constrained, albeit well-informed, model parameters. The minimum Lorentz factor and equipartition factor of the electron population in the lobe cannot realistically be constrained for individual sources and also faces moderate uncertainty in the mean value across radio sources; systematic errors in the mean values assumed for these parameters are handled using the calibration constants $b_1$ and $b_2$ respectively. The particle content in the lobe (i.e. ratio of leptons to baryons) is also modified using the $b_1$ calibration constant. Meanwhile, although the ratio of the lobe and expansion surface pressures is constrained by numerical simulations (Equation \ref{chi}, above), model assumptions likely do not perfectly represent actual sources, leading to potential large errors in the thinnest sources (i.e. $\chi \propto A^{-1.5}$ for $\beta \sim 1$); systematic errors in the exponent on the axis ratio are handled using the calibration constant $b_3$. Finally, the gas density at the radius of the lobe is modelled using density profiles for observed clusters and cosmological simulations (see Section \ref{sec:Cosmological environments}), however this modelling is subject both to small errors in the absolute scaling of the gas density and the cosmic evolution of the cluster mass function; these systematic errors are handled using calibration constants $b_2$ (shared with equipartition factor) and $b_4$ in the form $b_2 [1 + z]^{b_4}$. The four calibration constants are assumed to be zero by default (i.e. no correction required), however we discuss the use of these parameters in Section \ref{sec:APPLICATION TO HIGH-REDSHIFT SOURCES} when examining a sample of observed high-redshift sources.

\subsection{Cosmological environments}
\label{sec:Cosmological environments}

%The cluster environment inhabited by the lobed AGNs cannot be readily measured using radio-frequency observations (though polarimetry may provide a constraint in the future). The density profile is therefore simulated based on the distribution of cluster masses in semi-analytic galaxy evolution models \citep{Croton+2016}. 

The semi-analytic galaxy evolution \citep[SAGE;][]{Croton+2016, Raouf+2017} model \citep[update of original model;][]{Croton+2006} traces the evolutionary history of baryonic matter on top of existing large scale simulations of the dark matter. In this work we use the Bolshoi model \citep{Klypin+2011} to construct the dark matter framework for SAGE. This simulation traces galaxies in a box of side length $250\rm\, Mpc$$/h$ through cosmic time with outputs at regular intervals in redshift. The Bolshoi dark matter simulation assumes a WMAP5 cosmology with $\Omega_{\rm m} = 0.27$, $\Omega_\Lambda = 0.73$, $\Omega_{\rm b} = 0.0469$, $\sigma_8 = 0.82$, $h = 0.70$ and $n=0.95$; {the products we derive from this simulation are quite insensitive to small changes to the cosmological parameters (see Section \ref{sec:Hubble constant tension})}. We process Bolshoi simulation outputs at 15 redshifts up to $z = 5$ using SAGE to obtain mock galaxy populations across cosmic time. The \emph{RAiSE\textcolor{purple}{Red}} code extrapolates empirical relationships derived for $0<z<5$ beyond this maximum redshift to provide a good approximation of the distance to ultra-high redshift sources; however, in this work we conservatively limit ourseleves to objects below $z = 6$.
Based on the local 200\,MHz luminosity function \citep{Franzen+2020}, and assuming a luminosity evolution of $\sim$$(1+z)^{4\pm0.5}$ \citep{Seymour+2020}, the Bolshoi simulation volume is expected to include $0.7_{-0.4}^{+0.8}$ powerful radio galaxies with luminosities $L_{200\rm\,MHz} > 10^{28}\rm\, W\, Hz^{-1}$. The empirical relationships (discussed below) will therefore be derived, or more likely validated, using at least one example of even the most massive host galaxies expected for radio AGNs.

The mock galaxy populations based on the Bolshoi simulation are used to inform both the cluster mass function and the gas fraction as a function of redshift. Following \citet{Turner+2020}, we assume the low-redshift mass function of \citet{Girardi+2000} who find that a common Schechter function describes both galaxy groups and clusters. The Bolshoi simulations are used to extend their observations to higher redshifts by finding that the mass of the break in the Schechter function scales with redshift as approximately $(1 + z)^{-3}$. The relative normalisation of the cluster mass function between redshifts is not important in this work. The cluster mass function is further weighted by the AGN duty cycle \citep{Pope+2012}; high mass black holes and galaxies are known to have an enhanced probability of hosting AGNs compared to their lower mass counterparts \citep[e.g.][]{Sabater+2019}. Meanwhile, the cosmological evolution of the gas fraction as a function of halo mass is found to be well described by $f_{\rm gas} = 10^{w_0(z) - 14w_1} {(M_{\rm halo}/\rm M_\odot)}^{w_1}$, where $M_{\rm halo}$ is the simulated dark matter halo mass taken from SAGE, and $w_0(z) = \text{max}\{ -0.88 -0.03z, -0.92 + 0.001z \}$ and $w_1 = 0.05$ describe the redshift dependence. These gas fractions are further scaled by a small, constant factor based on low-redshift observations \citep{McGaugh+2010, Gonzalez+2013}. The spread in the distribution of gas fractions is approximately constant in log-space for all halo masses, but depends on redshift as $\delta f_{\rm gas} = 0.05 - 0.002z\,\rm dex$.

The shape of the gas density profiles is based on the \citet{Vikhlinin+2006} cluster observations following the method described in \citet{Turner+2015}\footnote{The gas density profile of the host galaxy can be neglected since core-confined radio AGNs (that are significantly impacted by their host galaxies) are excluded based on their `optically-thick' low-frequency spectral turnover (i.e. due to free-free or synchrotron self-absorption).}. The same shape profile can describe the environments of clusters with very different masses by scaling with the core density and virial radius of the halo. The virial radius is directly related to the mass of its halo through $r_{\rm halo} =  (G M_{\rm halo}/[100 H^2(z)])^{1/3}$, where $H(z)$ is the Hubble constant at redshift $z$, and $G$ is Newton's gravitation constant. %The \citet{Vikhlinin+2006} gas density profile is parameterised in terms of the virial radius and the core density, which in this work is arbitrarily taken as the density $\rho$ at the end of the lobe of radius $a$. 
The \citet{Vikhlinin+2006} gas density profile is integrated over the volume of the cluster within the virial radius and this unscaled mass is compared with the gas mass simulated using SAGE (based on the halo mass and gas fraction) to derive the density $\rho$ at the end of the lobe of radius $a$. The shape of the density profile is approximated locally using a power law of the form $\rho_{\rm gas}(r) = \rho\:\! [r/a]^{-\beta}$ to enable analytic modelling of the radio source evolution.

\subsection{Redshift probability density functions}

\subsubsection{Bayesian inference}

Probability density functions for the radio continuum redshift of a given source are derived using Bayesian statistics informed by a Monte Carlo simulation over the parameter space. The five attributes ($S_\nu$, $\theta$, $A$, $s$ and $\nu_{\rm b}$) are randomly sampled from within their uncertainty distributions in each of the Monte Carlo realisations. The three parameters describing the gas density profile of the cosmological environments ($a$, $\rho$ and $\beta$) are also randomly sampled based on typical cluster density profiles, the group and cluster mass function at the trial redshift $z$, and the angular size of the source at that redshift and realisation of the observables (Section \ref{sec:Cosmological environments}). The other model parameters, such as the equipartition factor and minimum Lorentz factor, take a fixed value across all realisations in the Monte Carlo simulation; these are corrected as necessary using the calibration constants.

The probability of a given radio AGN being located at a trial redshift $z$ given the measurement uncertainty distributions for its five attributes is given by:
\begin{gather}
\begin{split}
p(z \;\!|\;\! S_\nu, \theta, A, s, \nu_{\rm b}) &= \frac{1}{n} \sum_{i = 1}^{n}  \Big[1 - F(x(z \;\!|\;\! S_\nu, \theta, A, s, \nu_{\rm b}, i); k) \Big]\\
&\approx \frac{1}{n} \sum_{i = 1}^{n} e^{-x(z \;\!|\;\! S_\nu, \theta, A, s, \nu_{\rm b}, i)/2} ,
\label{prob}
\end{split}
\raisetag{20pt}
\end{gather}
where the summation, $i$, is over the $n$ Monte Carlo realisations with random variation in the observable and cosmological environment, and $F(x; k)$ is the cumulative chi-squared distribution for $k$ degrees of freedom. This distribution is exactly represented by an exponential function for $k=2$, and thus we assume $k \approx 2$ for computational efficiency in our calculation.
The chi-squared statistic is defined as:
\begin{equation}
\begin{split}
&x(z \;\!|\;\! S_\nu, \theta, A, s, \nu_{\rm b}, i) \\
&\quad\quad = \left[ \frac{d_{\rm M}(z \;\!|\;\! S_\nu, \theta, A, s, \nu_{\rm b}, i) - d_{{\rm M}\:\! (H_0, \Omega_{\rm m})}(z)}{\sigma_{d_{{\rm M}\:\! (H_0, \Omega_{\rm m})}(z)}} \right]^2 ,
\label{chisquared}
\end{split}
\end{equation}
where $d_{\rm M}(z \;\!|\;\! S_\nu, \theta, A, s, \nu_{\rm b}, i)$ is the transverse comoving distance given by Equation \ref{dM} at the trial redshift $z$ for the $i$-th Monte Carlo realisation of the cosmological environment and the value of the observed attributes within their uncertainty distributions. Meanwhile, $d_{{\rm M}\:\! (H_0, \Omega_{\rm m})}(z)$ is the expected transverse comoving distance at that redshift for the concordance cosmological model, and $\sigma_{d_{{\rm M}\:\! (H_0, \Omega_{\rm m})}(z)}$ is the standard deviation in that measurement due to uncertainites in the Hubble constant and matter density, assuming a flat universe.

\subsubsection{Prior probability density functions}

The prior probability density functions for the majority of parameters are informed by their measurement uncertainty distribution, whilst the cosmological environments are based on simulations and direct observations of the AGN duty cycle and the group and cluster mass function (Section \ref{sec:Cosmological environments}). However, we apply further constraints on these prior probability density functions by considering the Malmquist bias, imposing the speed of light as a conservative limit on the expansion speed of AGN lobes, and restricting jet kinetic powers to the broad range, $10^{35} \leqslant Q \leqslant 10^{45} \rm\, W\, Hz^{-1}$, expected for lobed radio sources. The expansion speed and jet kinetic power are derived from the observables in a given Monte Carlo realisation using the dynamical equations of \citet{Turner+2015}.

\subsubsection{Fourier filtered distribution}

\begin{figure}
\begin{center}
\includegraphics[width=0.47\textwidth,trim={15 15 45 40},clip]{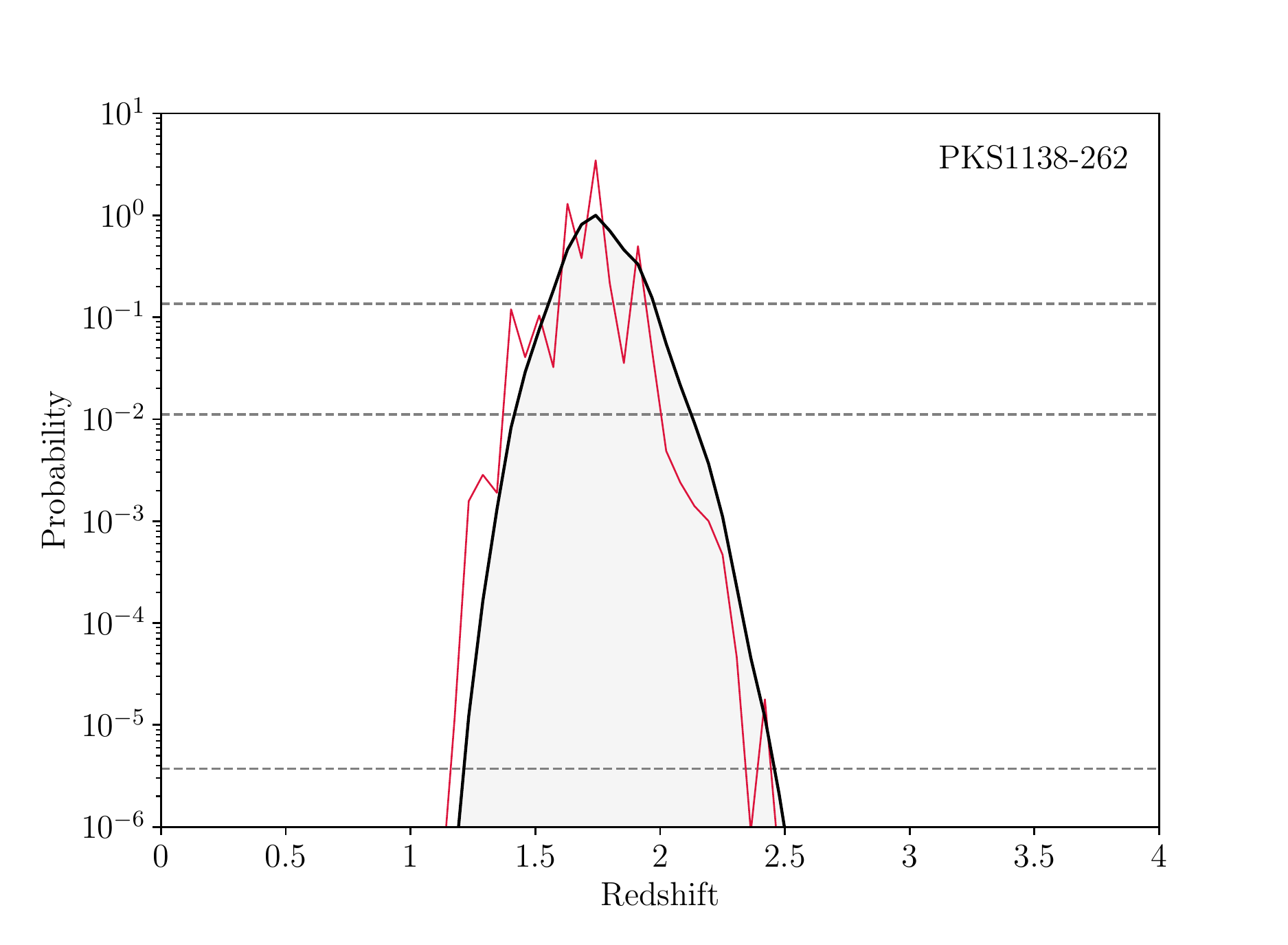}
\end{center}
\caption{Redshift probability density function for PKS\,1138$-$262 (see Section \ref{sec:APPLICATION TO HIGH-REDSHIFT SOURCES}), arbitrarily scaled to a maximum of unity. The thin red line plots the raw probability density function, $p(z)$, whilst the solid black line shows the Fourier filtered probability density function, $\tilde{p}(z)$. The dashed grey lines mark the approximate relative probability (based on a normal distribution) of the 2, 3 and 5$\sigma$ tails of the redshift probability density function. This graph is produced using the \texttt{RAiSERed\_plot()} function.}
\label{fig:probdist}
\end{figure}

The raw probability density function ($p(z)$; Equation \ref{prob}) includes ubiquitous high-frequency noise when sampling the parameter space with computationally practical numbers of Monte Carlo realisations (see Figure~\ref{fig:probdist}). {The simulations presented in this work use an adaptive number of realisations based on the strength (i.e. absolute probability) of the detected `signal', but capped at 100\,000 realisations per trial redshift. The number of realisations can be increased in the \emph{RAiSE\textcolor{purple}{Red}} code to reduce the noise level by a factor of $1/\sqrt{n}$ but increase computation time by a factor of $n$.}
The `signal' noise must be suppressed as it not only prevents the peak of the distribution from being correctly identified, but in sources with a relatively low probability of matching any set of parameters, random noise spikes (resulting from just one or two realisations) can appear in any redshift bin by chance.
The high-frequency noise is removed by applying a real-valued Fast Fourier Transform to the probability density function, $p(z)$. The transformed function, $P(1/z)$, is convolved with a decaying exponential function to suppress the amplitude of the high-frequency components in the Fourier spectrum. The noise-filtered probability density function, $\tilde{p}(z)$, is obtained by taking the inverse Fourier Transform of the convolved function. The maximum amplitude of this Fourier filtered probability density function is arbitrarily scaled to unity for ease of comparison between sources.% and used to derive the mean and standard deviation of the redshift distribution.

The radio continuum redshift for a given lobe is mathematically defined as the mean of the approximately normally distributed Fourier filtered probability density function; i.e. $z^* = \int z \tilde{p}(z) dz / \int \tilde{p}(z) dz$. The uncertainty in the redshift is taken as the standard deviation of the probability density function. We plot the shape of the filtered probability density function where practicable throughout this work in addition to quoting summary statistics. The probability density functions for the two lobes of a given source can also be combined to yield a single robust estimate of the photometric redshift.

\subsection{Spectroscopic calibration}

The calibration constants in the \emph{RAiSE\textcolor{purple}{Red}} model described in the previous sections can be constrained using observations of at least five\footnote{There are four calibration constants so an absolute minimum of five independent measurements are required.} radio AGNs with known spectroscopic redshifts. The sum of squared differences between the radio continuum redshifts (i.e. mean of the probability density function) and the spectroscopic redshifts for the calibration sample is minimised to find the optimal values for $b_1$, $b_2$, $b_3$ and $b_4$. The optimisation is performed using the \texttt{noisyopt} package in \emph{python}, a robust pattern search algorithm with an adaptive number of function evaluations \citep{Mayer+2016}. Importantly, this algorithm does not use the function derivative to locate the minimum, an approach that would not be viable in this work given random noise is more significant that the gradient for small perturbations in the calibration constants. Despite the greatly improved computational efficiency of this algorithm over a brute force technique the computational time for a sample of 15 objects is approximately 4-6 hours on a typical laptop. The calibration of the \emph{RAiSE\textcolor{purple}{Red}} model for use in large sky surveys will likely need to use supercomputer time or only a small subsample of calibrators.

\section{ASSESSMENT OF METHOD ON MOCK SOURCES}
\label{sec:ASSESSMENT OF METHOD ON MOCK SOURCES}

\subsection{Simulation of Cygnus A-like population}

Cygnus A is a low-redshift radio source \citep[spectroscopic redshift of $z = 0.056075 \pm 0.000067$;][]{Owen+1997} with a double FR-II lobe morphology. \citet{Steenbrugge+2010} reported core and hotspot removed measurements of the flux density in the east and west lobes of Cygnus A at six frequencies from $151\rm\, MHz$ to $15\rm\, GHz$. The size and axis ratio of the two lobes are measured from the $5\rm\, GHz$ radio images following \citet{Turner+2019}; the uncertainty in the length of each lobe is taken as the size of the synthesised beam at this frequency \citep[$\theta_{\rm res}$;][]{Carilli+1996}. The properties of the electron population in each lobe are constrained by fitting the radio spectra using the continuous injection (CI) model following the method of \citet{Turner+2018b}. The $151\rm\, MHz$ to $15\rm\, GHz$ flux density measurements are supplemented by $74\rm\, MHz$ observations from \citet{Cohen+2007} to better constrain the electron energy injection index, $s$; the uncertainties on the \citet{Steenbrugge+2010} measurements are estimated following \citet{Carilli+1996}. The radio continuum observations for the two lobes of Cygnus A are summarised in Table~\ref{tab:CygnusA}.%; i.e. the $151\rm\, MHz$ flux density ($S_{151\rm\, MHz}$), lobe length ($D$), axis ratio ($A$), electron energy injection index ($s$), and the break frequency ($\nu_{\rm b}$).

\begin{table*}
\begin{center}
\caption[]{Radio continuum attributes for the two lobes of Cygnus A based on the multi-wavelength study of \citet{Steenbrugge+2010}. The second through sixth columns list: the flux density of each lobe; lobe length; axis ratio; electron energy injection index; and the break frequency. Measurement uncertainties are quoted at the 1$\sigma$ level (unless otherwise described in the text); the uncertainties for $s$ and $\nu_{\rm b}$ presented here do not consider any source of systematic error (e.g. flux density variations from inhomogenous magnetic fields).}
\label{tab:CygnusA}
\renewcommand{\arraystretch}{1.1}
\setlength{\tabcolsep}{8pt}
\begin{tabular}{ccccccc}
\hline\hline
\multirow{2}{*}{Source}&\multicolumn{5}{c}{Radio continuum attributes} \\
&$S_{151\rm\, MHz}$ (Jy)&$\theta$ (arcsec)&$A$&$s$&$\nu_{\rm b}$ (log Hz) \\
\hline
Cygnus\,A East&5960$\pm$450&58.6$\pm$0.4&2.8&2.485$\pm$0.009&9.243$\pm$0.017 \\
Cygnus\,A West&4750$\pm$350&67.3$\pm$0.4&3.0&2.436$\pm$0.008&9.305$\pm$0.015 \\
\hline
\end{tabular}
\end{center}
\end{table*}

The \emph{RAiSE\textcolor{purple}{Red}} model is tested for radio AGNs across redshifts $z = 0$ to 6 by shifting the observed radio continuum observations of Cygnus A to higher redshifts. The flux density and lobe angular size are converted for the cosmology assumed in this work, whilst intrinsic properties of the source (axis ratio and electron energy injection index) are unchanged. The observer-frame break frequency measured for an identical source to Cygnus A (at redshift $z_0$) but located at redshift $z$ is found from rearranging Equation 9 of \citet{Turner+2019},
\begin{equation}
\nu_{\rm b}(z) = \frac{1 + z_0}{1 + z} \left[\frac{B^2/({0.318\rm\, nT})^2 + (1 + z_0)^4}{B^2/({0.318\rm\, nT})^2 + (1 + z)^4} \right]^2 \nu_{\rm b}(z_0) ,
\label{spectral}
\end{equation}
where $B$ is the magnetic field in the lobe. The magnetic field strength is constrained using the dynamical model in Equation \ref{Bfield} for each lobe to provide an estimate calibrated for any systematic errors in the model, based on the obsevred properties of Cygnus A at $z = 0.056$. %We note that the exact value used here are not that important, however we choose sensible values to ensure the simulated Cygnus A-like sources should be somewhat representative of the actual high-redshift population.

\begin{figure}
\begin{center}
\includegraphics[width=0.42\textwidth,trim={25 20 50 45},clip]{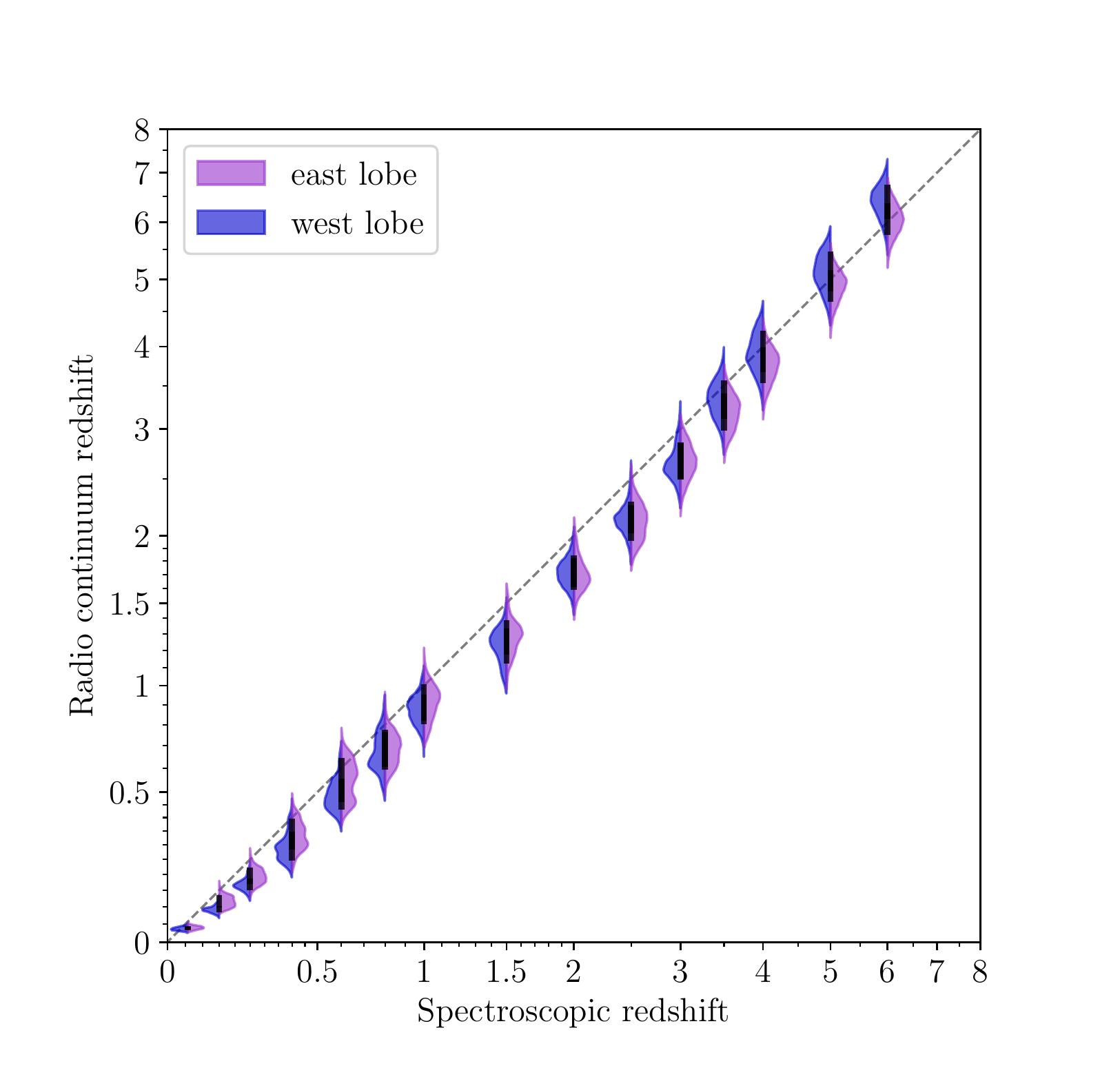}
\end{center}
\caption{Radio continuum redshifts estimated for two lobes of the mock Cygnus A-like population as a function of their true redshifts. The violin plot shows the probability density function for the east (purple) and west (blue) lobes for each simulated source, with the 1$\sigma$ confidence level shown by a narrow black rectangle in the center of the `violin'. The probability density functions are trunctated at the 3$\sigma$ level for clarity. The one-to-one line is shown on the plot as a grey dashed line. This graph was produced using the \texttt{RAiSERed\_plotzz()} function.}
\label{fig:CygnusA}
\end{figure}

The radio continuum attributes of the east and west lobes of Cygnus A are shifted from $z = 0.056$ to progressively higher redshifts up to $z = 6$. %higher redshifts of $z = 0.15$, 0.25, 0.4, 0.6, 0.8, 1, 1.5, 2, 2.5, 3, 3.5, 4, 5, and 6; i.e. a total of 15 redshifts are considered. %The cosmological environments assumed in our model are expected to remain valid for the highest redshift bins noting the simulated gas fraction is constant for $z\geqslant2$ when AGN feedback is not significant.
The cosmological environments assumed in our model are not strictly valid for the Cygnus A-like sources shifted to substanially higher redshifts (i.e. the observed properties are influenced by the actual $z\sim0.056$ environment). However, in order to maintain the correct model response to small variations in redshift we choose to use the cosmological environment appropriate to each trial redshift \footnote{The findings in this section are unchanged if we assume the $z=0.056$ environment at all redshifts, although the model becomes less stable around redshifts $z=1.5-2$, coinciding with the turnover in the angular diameter distance--redshift function.}. The best estimate radio continuum redshifts for the east and west lobes of these Cygnus A-like sources are shown in Figure~\ref{fig:CygnusA} as a function of their true redshifts (i.e. shifted spectroscopic redshifts). The radio continuum redshifts are consistent with the true redshifts within the 1$\sigma$ measurement uncertainties for the majority of the simulated Cygnus A-like sources, except those at $z = 0.15$-0.6 and from $z=1$-2.5 (these are consistent at the 2$\sigma$ level). We note that the absolute scaling need not be correct at this stage since our model includes calibration terms (i.e. $b_1$, $b_2$, $b_3$ and $b_4$); however, this agreement supports both our modelling of the cosmological environments and the (uncalibrated) values chosen for the less well constrained model parameters ($q$, $\gamma$, $\mathcal{Y}$).

\subsection{Sensitivity of estimates to measurement uncertainties}

In this section, the simulated population of Cygnus A-like sources is used to test the ability of the \emph{RAiSE\textcolor{purple}{Red}} model to estimate redshifts for unresolved sources (i.e. size upper limits) and objects with poorly constrained break frequencies. The other radio continuum observables have smaller (absolute) exponents in the distance measure equation \citep{Turner+2019} and thus modest uncertainties are not expected to adversely effect the redshift estimates.

\subsubsection{Unresolved sources}

\begin{figure}
\begin{center}
\includegraphics[width=0.42\textwidth,trim={25 20 50 45},clip]{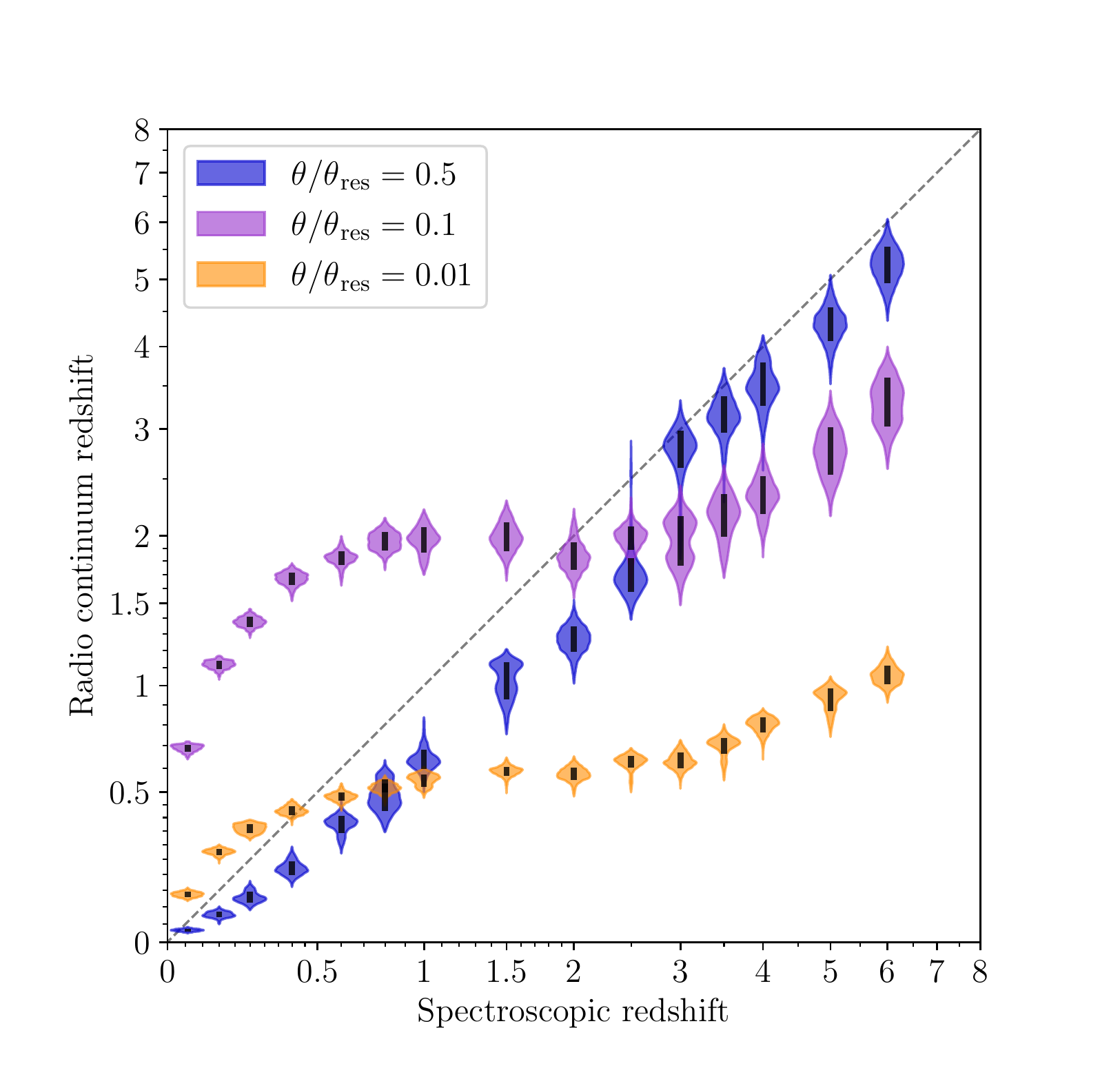}
\end{center}
\caption{Radio continuum redshifts estimated for an unresolved mock Cygnus A-like population as a function of their true redshifts. The violin plot shows the probability density function for the east lobe assuming its angular size is a factor two (blue), ten (purple) or one-hundred (orange) lower than the size upper limit imposed by the survey resolution. See the caption of Figure~\ref{fig:CygnusA} for a complete description of the plot.}
\label{fig:CygnusA_size}
\end{figure}

The \emph{RAiSE\textcolor{purple}{Red}} code (included in the online supplementary material) can take upper limits, or any asymmetric uncertainties, as inputs for any of the five radio continuum attributes. These are specified as a skewed normal distribution with location (e.g. mean), $\xi$; scale (e.g. standard deviation), $\omega$; and skewness, $\lambda$. The skewness is set to $\lambda \rightarrow -\infty$ for an upper limit and $\lambda \rightarrow \infty$ for a lower limit; this results in a uniform prior probability density function for permitted values. We test our code's performance on unresolved sources by assuming that the simulated lobe length of the Cygnus A-like sources, $\theta$, is a factor of two, ten or one-hundred lower than the size upper limit, $\theta_{\rm res}$. The best estimate redshifts for this mock population of sources are shown in Figure~\ref{fig:CygnusA_size}. The radio continuum redshifts broadly agree with the spectroscopic redshifts if the angular size of the lobe is half the survey resolution limit, and are strongly correlated at high-redshift ($z \geqslant 2$) for sizes a factor of ten below the resolution limit. For the most unresolved sources with $\theta/\theta_{\rm res} < 0.01$, the radio AGN dynamics are too poorly constrained to yield accurate redshift estimates. The estimated radio continuum redshifts in this instance are always less than $z \leqslant 1$, so despite their ineffectiveness, they only corrupt a small region of the posterior probability density function. Size upper bounds are therefore expected to be accurate for unresolved sources with estimated radio continuum redshifts $z \geqslant 2$. Size upper bounds are not viable at lower redshifts unless the lobe size is comparable to the angular survey resolution limit. 

It may be possible to constrain the degeneracy in the model in these situations through independent arguments or measurements. In particular, as the angular diameter distance scale peaks at $8.7\rm\, kpc/arcsec$ the lobes of any extended radio AGN that has expanded beyond the galaxy cannot be much less than an arcsecond in angular size (cf. typical survey resolutions of 0.2-1.5$\,\rm arcsec$). Radio AGNs confined to their host galaxies can be excluded by an `optically-thick' low-frequency turnover (i.e. due to free-free or synchrotron self-absorption), thus providing a lower bound on the angular size (such that $\theta/\theta_{\rm res} \gtrsim 0.5$). In this manner, it may be possible to accurately constrain the radio continuum redshifts of unresolved sources at any redshift, if a sufficiently high resolution survey is employed.

\subsubsection{Poorly constrained spectral breaks}

\begin{figure}
\begin{center}
\includegraphics[width=0.42\textwidth,trim={25 20 50 45},clip]{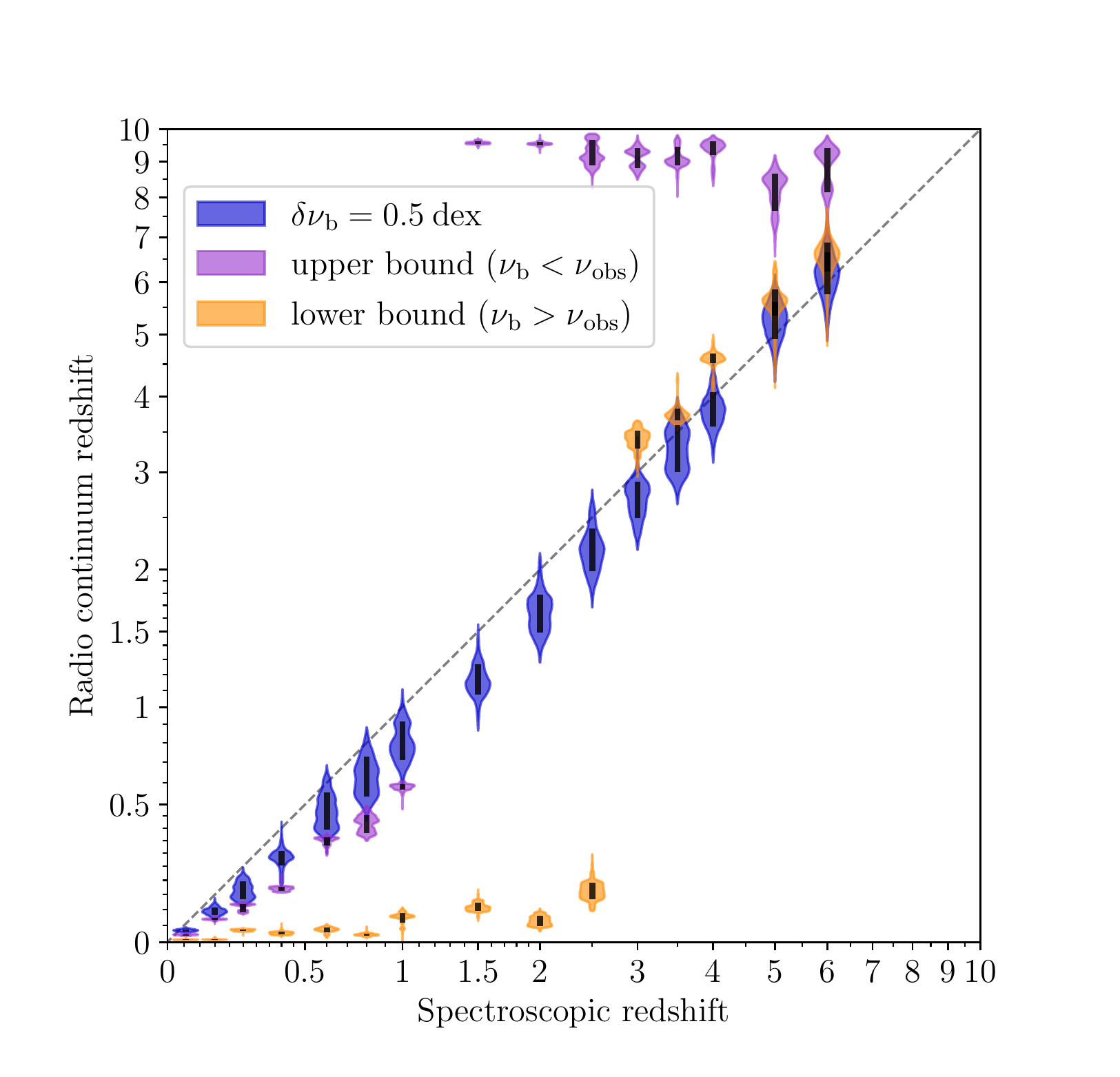}
\end{center}
\caption{Radio continuum redshifts estimated for mock Cygnus A-like sources with poorly constrained break frequencies as a function of their true redshifts. The violin plot shows the probability density function for the east lobe assuming either: the break frequency is detected but has a large measurement uncertainty (blue--scenario 1); the break frequency is not detected but must be at lower frequencies (purple--scenario 2); or the break frequency is not detected but must be at higher frequencies (orange--scenario 3). See the caption of Figure~\ref{fig:CygnusA} for a complete description of the plot.}
\label{fig:CygnusA_break}
\end{figure}

The sensitivity of the radio continuum redshift estimates to the break frequency is similarly tested by considering three scenarios: (1) the break frequency is detected but has a large measurement uncertainty; (2) the break frequency is not detected but must be lower than the observed frequencies (i.e. aged spectrum; $\alpha \gtrsim 1$); and (3) the break frequency is not detected but must be at higher frequencies (i.e. freshly injected spectrum; $\alpha \lesssim 1$). The first of these options is modelled for the Cygnus A-like sources by assuming an uncertainty of $0.5\rm\, dex$. In the latter two scenarios, an upper bound is placed a factor of ten below the simulated break frequency or a lower bound a factor of ten above the break frequency respectively; the prior probability density functions are assumed to be uniform in log-space. The radio continuum redshifts estimated for the Cygnus A-like population are shown in Figure~\ref{fig:CygnusA_break} for each of the three scenarios for the break frequency. 

Redshifts estimated using \emph{RAiSE\textcolor{purple}{Red}} remain in broad agreement with the spectroscopic redshifts despite a large measurement uncertainty in the break frequency, albeit with the correct mean value (scenario 1). %, however a couple of mock sources from $z = 0.15$\,-\:\!2 are now only consistent at the looser 2$\sigma$ confidence level. 
Mock radio AGNs that exhibit an aged spectrum (scenario 2) have radio continuum redshifts $z \sim 10$ when their true redshift is $z > 1$, but are consistent with their spectroscopic counterparts for $z \leqslant 1$. That is, the probability density functions are corrupted by noise values at the maximum redshift of our code if the source is not at low-redshift. By contrast, AGNs showing a freshly injected spectrum at all observed frequencies (scenario 3) accurately have their radio continuum estimated if their true redshift is $z > 3$, but are inconsistent with their spectroscopic counterparts at lower redshifts. 
%can similarly be used to classify sources with upper bounds on the break frequency into low- or high-redshift. 
For these two scenarios, the \emph{RAiSE\textcolor{purple}{Red}} model can therefore gauge whether such objects are at low- or high-redshift (i.e. does it corrupt the probability density function or not), in addition to small redshift ranges with good accuracy for both the lower and upper bounds. 
The spectral shape of target radio AGNs must therefore be constrained reasonably confidently to ensure an unique and precise radio continuum redshift is fitted, however a limiting value does not preclude a broad categorisation as a $z \leqslant 1$ or $z > 3$ object.

\subsubsection{Hubble constant tension}
\label{sec:Hubble constant tension}

\begin{figure}
\begin{center}
\includegraphics[width=0.42\textwidth,trim={25 20 50 45},clip]{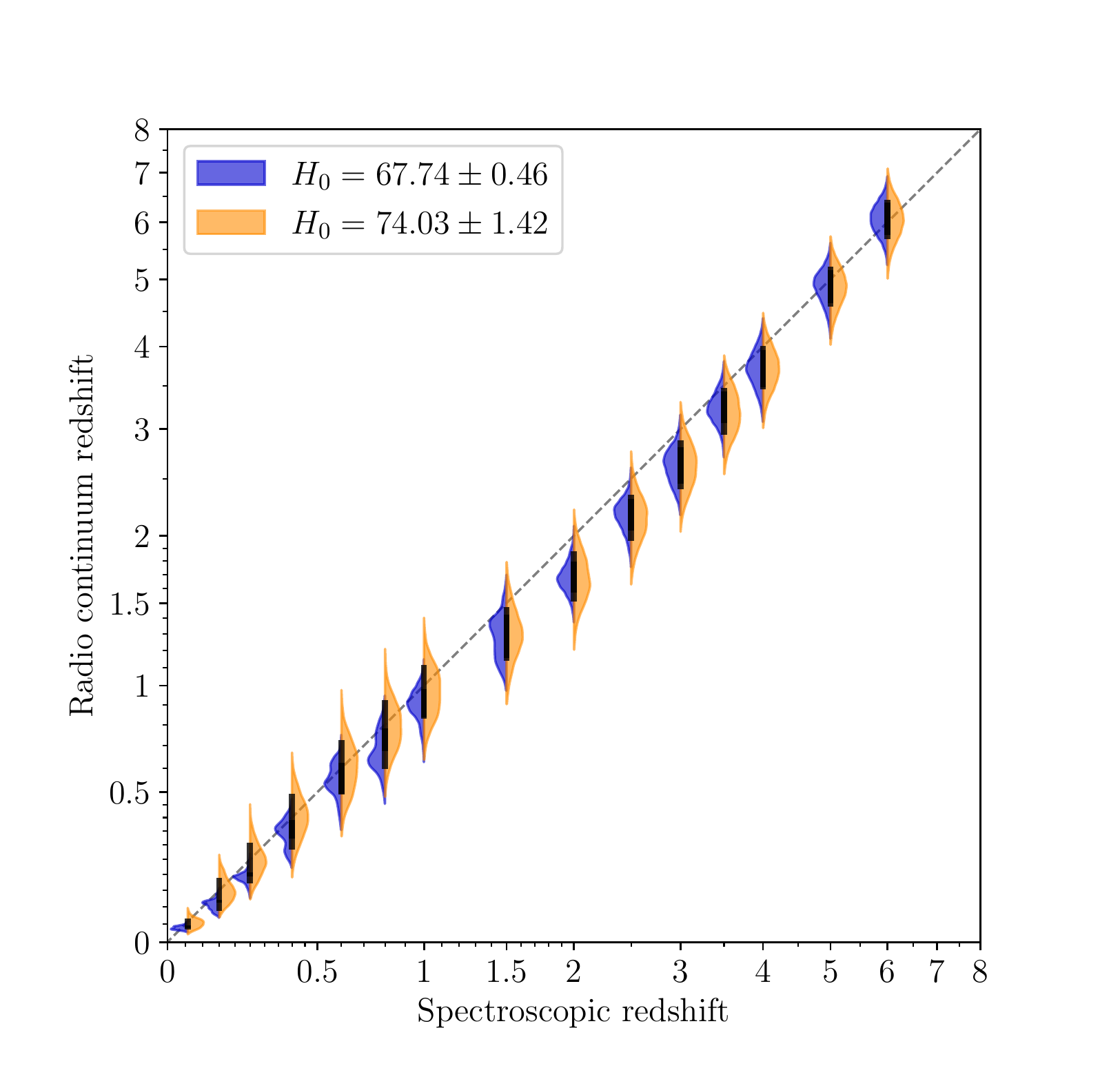}
\end{center}
\caption{{Radio continuum redshifts estimated for mock Cygnus A-like sources for different values of the Hubble constant. The violin plot shows the probability density function for the east lobe assuming either: the Hubble constant measured from the cosmic microwave background (blue); or the local value measured using supernovae and Cepheid variables (orange). See the caption of Figure~\ref{fig:CygnusA} for a complete description of the plot.}}
\label{fig:CygnusA_hubble}
\end{figure}

{Hubble constant measurements using the cosmic microwave background are known to be in tension with Type-1a supernovae and Cepheid variables based estimates \citep[e.g.][]{Planck+2016, Riess+2016, Riess+2019, Freedman+2019}. This tension is not considered in the statistical uncertainties for the cosmological distances assumed in our Bayesian inference, in particular the distance for each trial redshift in the chi-squared statisitic and the scaling of the density profile of their cosmological environments. We test the sensitivity of the radio continuum redshift estimates to the value of the Hubble constant by considering a current cosmic microwave background measurement \citep[][$H_0 = 67.74\pm 0.46\rm\, km\,s^{-1}\, Mpc^{-1}$]{Planck+2016} and one of the more discrepant local estimates based on Cepheid variables in the Large Magellanic Cloud \citep[][$H_0 = 74.03\pm 1.42\rm\, km\,s^{-1}\, Mpc^{-1}$]{Riess+2019}. The radio continuum redshifts estimated for the Cygnus A-like population are shown in Figure~\ref{fig:CygnusA_hubble} for both values of the Hubble constant. The radio continuum redshifts are larger for mock sources with true redshifts $z \leqslant 1$ when using the local value of the Hubble constant, though given their large uncertainties, these are still consistent with the cosmic microwave background based estimates. Mock sources with true redshifts $z > 1$ are unaffected by the modest change in the value of the Hubble constant. Redshifts estimated using the \emph{RAiSE\textcolor{purple}{Red}} model are therefore expected to be robust to the current uncertainty in the cosmological parameters. The arguments presented here also validate our earlier claim that modest differences between the Bolshoi dark matter simulation cosmology and that assumed in this work have an insignificant effect on our results; i.e. both positive and negative differences from the Hubble constant assumed in the dark matter simulation yield consistent redshift estimates.}

\section{APPLICATION TO HIGH-REDSHIFT SOURCES}
\label{sec:APPLICATION TO HIGH-REDSHIFT SOURCES}

The \emph{RAiSE\textcolor{purple}{Red}} model is now applied to a sample of observed sources across a broad range of redshifts to assess the viability of our technique in measuring redshifts for large-sky surveys. The error in the model due to uncaptured physical processes in extended radio sources is also estimated for our calibrated algorithm.

\subsection{Radio AGN samples}

\subsubsection{HeRGE radio sources}

The \emph{Herschel Radio Galaxy Evolution} project \citep[HeRGE;][]{Drouart+2014} is a comprehensive imaging survey of 70 radio galaxies at redshifts $1 < z < 5.2$. The host galaxy of each object has been identified and includes a spectroscopic redshift. The HeRGE project sample combines literature radio-frequency observations of the integrated lobe luminosity at observer-frame frequencies across the range 10\,MHz to 15\,GHz \citep[e.g.][full SEDs will be presented in \citealt{Drouart+2020}]{Carilli+1997, Pentericci+2000, DeBreuck+2010}. The literature flux density measurements of 34 HeRGE objects are supplemented by broadband \emph{Murchison Widefield Array} (MWA) observations from 72 to 231\,MHz; these observations are split into 20 subbands of 8\,MHz bandwidth \citep{Hurley-Walker+2017}. The properties of the electron population in this subsample are constrained by fitting the radio spectra using the CI model \citep[see e.g. Equations 1-3 and 8 of][]{Turner+2018b}; specifically, the electron energy injection index, $s$, and break frequency, $\nu_{\rm b}$, are optimised by minimising the relevant chi-squared statistic. Meanwhile, the angular size and axis ratio of the combined two lobes is measured from high-resolution \emph{Very Large Array} (VLA) images at 4.7-4.9 and 8.2-8.5\,GHz (see Table~\ref{tab:HeRGE} for image references). The uncertainty in the length of the combined two lobes is taken as the size of the synthesised beam (i.e. $\theta_{\rm res}$) as for Cygnus A, whilst the uncertainty in the axis ratio is based on the difference in the maximum width of the two lobes (i.e. near each end of the source). 

\begin{figure}
\begin{center}
\includegraphics[width=0.47\textwidth,trim={15 15 45 40},clip]{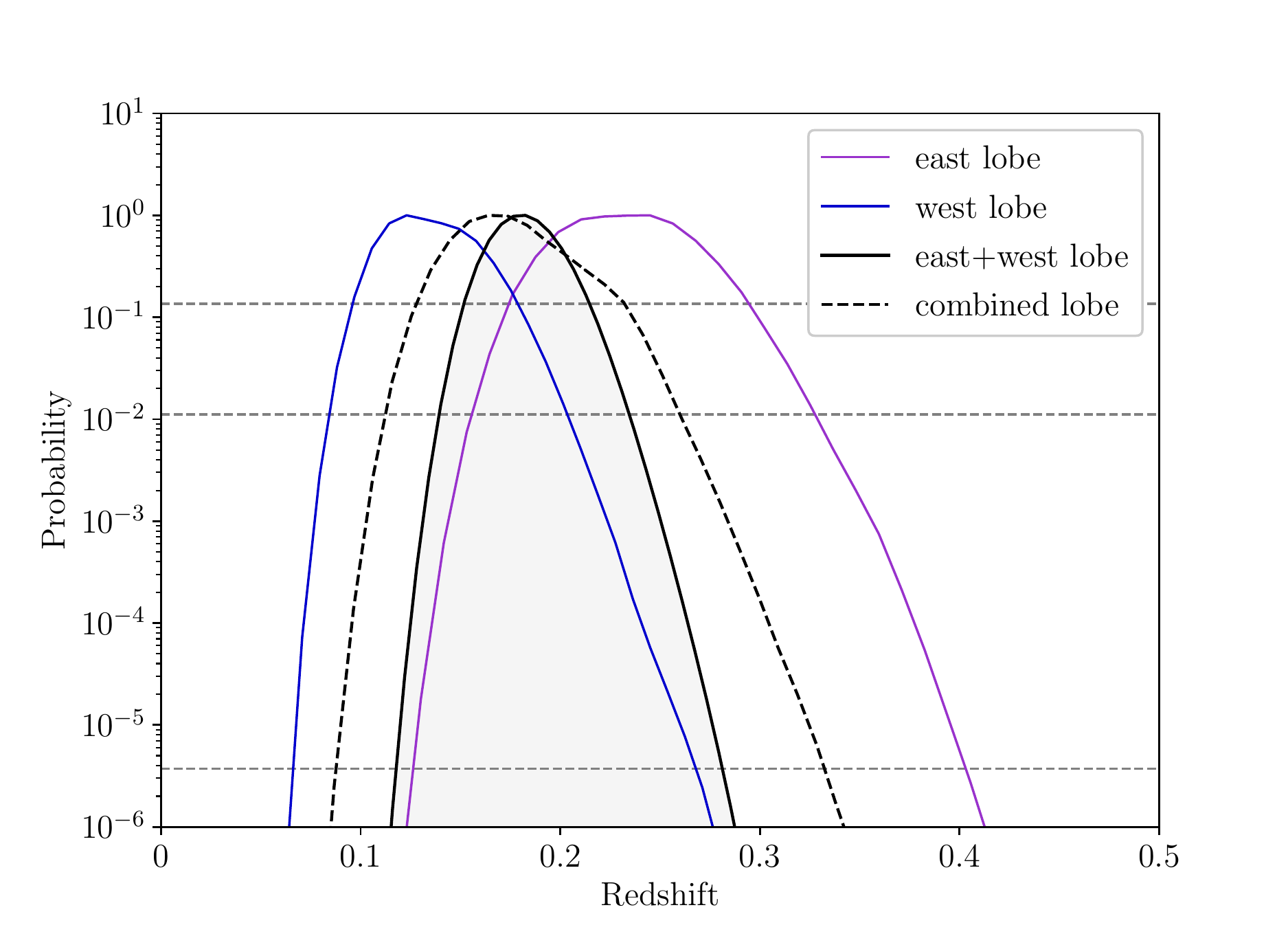}
\end{center}
\caption{{Redshift probability density function for the two lobes of 3C388. The Fourier filtered probability density function, $\tilde{p}(z)$, is shown for the east lobe (purple), the west lobe (blue), and a combined lobe (dashed black) which assumes an average value for each of the five radio continuum attributes across the two lobes. In comparison, the solid black line (and shading) shows the product of the redshift probability density functions for the east and west lobes; i.e. acknowledging that the two lobes must be at the same redshift. See the caption of Figure~\ref{fig:probdist} for a complete description of the plot.}}
\label{fig:combined_dist}
\end{figure}

{The two lobes are not considered separately because the location of the core is not apparent at radio frequencies for these high-redshift objects. However, we find the redshift probability density function for the combined two lobes of 3C388 (see Section \ref{sec:3C radio sources}) is consistent with the product of the probability density functions for the separate lobes, albeit the distribution is wider by $\sqrt 2$ (see Figure \ref{fig:combined_dist}). This agreement occurs despite the two lobes separately having highly discrepant radio continuum redshift estimates ($\Delta z^*/\bar{z}^* \sim 0.8$). The necessary use of a single set of radio continuum attributes across both lobes is therefore not expected to negatively affect the redshift estimates.}

\begin{table*}
\begin{center}
\caption[]{Radio continuum attributes of HeRGE sources with confident spectral fits. The source name and its host galaxy spectroscopic redshift are listed in the first two columns. The third through seventh columns list: the average flux density (half the source flux density); average lobe length (half the total source size); average axis ratio; the electron energy injection index; and the break frequency. The source of the high-resolution radio map is listed in the final column.}
\label{tab:HeRGE}
\renewcommand{\arraystretch}{1.1}
\setlength{\tabcolsep}{8pt}
\begin{tabular}{cccccccc}
\hline\hline
\multirow{2}{*}{Source}&\multirow{2}{*}{Redshift}&\multicolumn{5}{c}{Radio continuum attributes (average of two lobes)}&\multirow{2}{*}{Reference} \\
&&$S_{151\rm\, MHz}$ (Jy)&$\theta$ (arcsec)&$A$&$s$&$\nu_{\rm b}$ (log Hz) &
\\
\hline
PKS\,0529$-$549&2.57&2.78$\pm$0.22&0.6$\pm$0.33&$\geqslant\:\!$1.6&2.474$\pm$0.005&9.160$\pm$0.014&\citet{Broderick+2007} \\
PKS\,1138$-$262&2.15&5.67$\pm$0.45&7.9$\pm$0.125&5.3$\pm$0.1&2.887$\pm$0.008&9.169$\pm$0.015&\citet{Carilli+1997} \\
USS\,1243+036&3.57&2.23$\pm$0.18&3$\pm$0.115&6.9$\pm$0.8&2.739$\pm$0.007&9.005$\pm$0.014&\citet{van Ojik+1996} \\
USS\,1558$-$003&2.52&2.04$\pm$0.17&4.6$\pm$0.115&4.2$\pm$0.7&2.670$\pm$0.009&9.295$\pm$0.024&\citet{Pentericci+2000} \\
USS\,1707+105&2.34&1.44$\pm$0.12&11.25$\pm$0.115&$\leqslant\:\!$13.5&2.528$\pm$0.008&8.963$\pm$0.016&\citet{Pentericci+2001} \\
\hline
\end{tabular}
\end{center}
\end{table*}

The HeRGE sample is reduced to five high-quality candidates to test the efficacy of the \emph{RAiSE\textcolor{purple}{Red}} model at high-redshift. Specifically, we exclude sources that do not have an FR-II lobe morphology, and those whose spectrum is either consistent with a straight line or has an `optically-thick' low-frequency turnover (i.e. due to free-free or synchrotron self-absorption). The chosen sources further have fitted break frequencies that are robust to the removal of any two flux density measurements; this is not a necessary model requirement, rather we seek to assess our technique using only sources for which we are highly confident in the accuracy of the five observable model parameters.% (i.e. can distinguish errors in our technique and the preliminary interpretation of the observations). 

The selected sources, PKS\,0529$-$549, PKS\,1138$-$262, USS\,1243+036, USS\,1558$-$003 and USS\,1707+105, are located between redshift $z = 2.15$ and 2.57 with a single source (USS\,1243+036) at $z = 3.57$. Importantly, the observed attributes individually show no correlation with spectroscopic redshift; i.e. these objects show sufficient variation in intrinsic properties not to follow redshift--flux density or redshift--angular size relationships. Specifically, redshift explains only 15\% of the variation in the flux density measurements, 27\% in the angular size, $<$\,1\% in the injection index, and 13\% in the break frequency. Meanwhile, the axis ratio of two of the sources are only measured as an upper or lower bound; PKS\,0529$-$549 has an angular size within a factor of two of the beam size and only an upper bound can be placed on the width of the source, whilst the high-resolution images of USS\,1707+105 only show emission close to the hotspot (i.e. upper bound on the lobe width). The observed attributes derived from the multi-frequency radio images of the five HeRGE objects are summaried in Table~\ref{tab:HeRGE}.

\subsubsection{3C radio sources}
\label{sec:3C radio sources}

The \emph{Third Cambridge Catalogue of Radio Sources} (3CRR) is a complete sample of extragalactic radio sources in the Northern Hemishpere with 178\,MHz flux density $>$\,10.9\,Jy \citep{Laing+1983}. \citet{Mullin+2008} presented a catalogue comprising 98 low-redshift ($z < 1$) 3C sources with measurements of the flux density, angular size and axis ratio of each lobe. Following \citet{Turner+2018b}, we supplement the 178\,MHz flux density measurement with multi-frequency radio observations from \citet{Laing+1980}. The properties of the electron population are derived by fitting the CI model to the radio spectra for the integrated flux density arising from both lobes. We arbitrarily choose five 3C sources which have an FR-II morphology in both lobes and fitted break frequencies that are robust to the removal of two flux density measurements. The selected sources, 3C20, 3C219, 3C244.1, 3C388 and 3C438, have their observed properties summarised in Table~\ref{tab:3C} for both lobes. The two lobes are designated as east--west or north--south based on which axis the source is most closely aligned.

\begin{table*}
\begin{center}
\caption[]{Radio continuum attributes of \emph{Third Cambridge Catalogue of Radio Sources} (3C) sources with confident spectral fits. The columns are the same as for Table~\ref{tab:HeRGE}; the flux density, angular size and axis ratio radio continuum attributes are taken from \citet{Mullin+2008}.}
\label{tab:3C}
\renewcommand{\arraystretch}{1.1}
\setlength{\tabcolsep}{8pt}
\begin{tabular}{cccccccc}
\hline\hline
\multirow{2}{*}{Source}&\multirow{2}{*}{Redshift}&\multicolumn{5}{c}{Radio continuum attributes} \\
&&$S_{178\rm\, MHz}$ (Jy)&$\theta$ (arcsec)&$A$&$s$&$\nu_{\rm b}$ (log Hz)
\\
\hline
3C20 East&0.174&21.64$\pm$0.13&24.55$\pm$0.22&5.0&2.222$\pm$0.001&9.759$\pm$0.001 \\
3C20 West&0.174&25.12$\pm$0.11&25.91$\pm$0.22&3.7&2.222$\pm$0.001&9.759$\pm$0.001 \\
3C219 North&0.1744&21.10$\pm$1.06&81.82$\pm$1.40&3.2&2.439$\pm$0.004&9.622$\pm$0.014 \\
3C219 South&0.1744&23.80$\pm$1.19&99.09$\pm$1.40&7.0&2.439$\pm$0.004&9.622$\pm$0.014 \\
3C244.1 North&0.428&12.95$\pm$0.65&28.45$\pm$0.40&10.0&2.562$\pm$0.005&9.930$\pm$0.022 \\
3C244.1 South&0.428&9.18$\pm$0.46&25.09$\pm$0.40&10.5&2.562$\pm$0.005&9.930$\pm$0.022 \\
3C388 East&0.0908&12.06$\pm$0.60&22.1$\pm$0.8&4.6&2.318$\pm$0.004&9.723$\pm$0.015 \\
3C388 West&0.0908&14.75$\pm$0.74&20.5$\pm$0.8&2.9&2.318$\pm$0.004&9.723$\pm$0.015 \\
3C438 North&0.290&23.44$\pm$1.17&12.86$\pm$0.23&2.9&2.450$\pm$0.003&9.106$\pm$0.006 \\
3C438 South&0.290&25.29$\pm$1.26&12.00$\pm$0.23&2.8&2.450$\pm$0.003&9.106$\pm$0.006 \\
\hline
\end{tabular}
\end{center}
\end{table*}

\subsection{Calibration and systematic error estimation}

\begin{figure*}
\begin{center}
\includegraphics[width=0.42\textwidth,trim={25 20 55 45},clip]{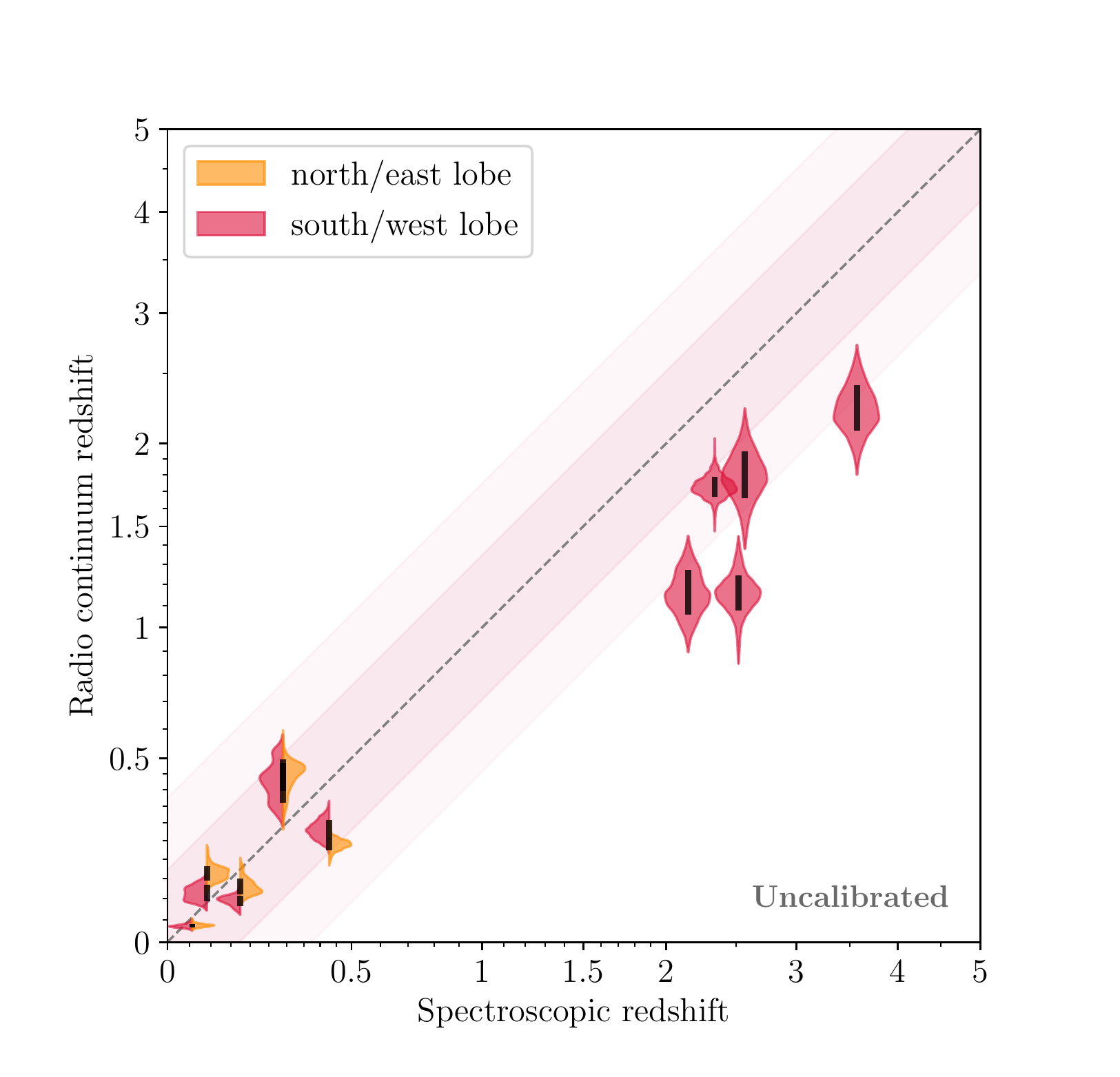}\quad\quad\quad\includegraphics[width=0.42\textwidth,trim={25 20 55 45},clip]{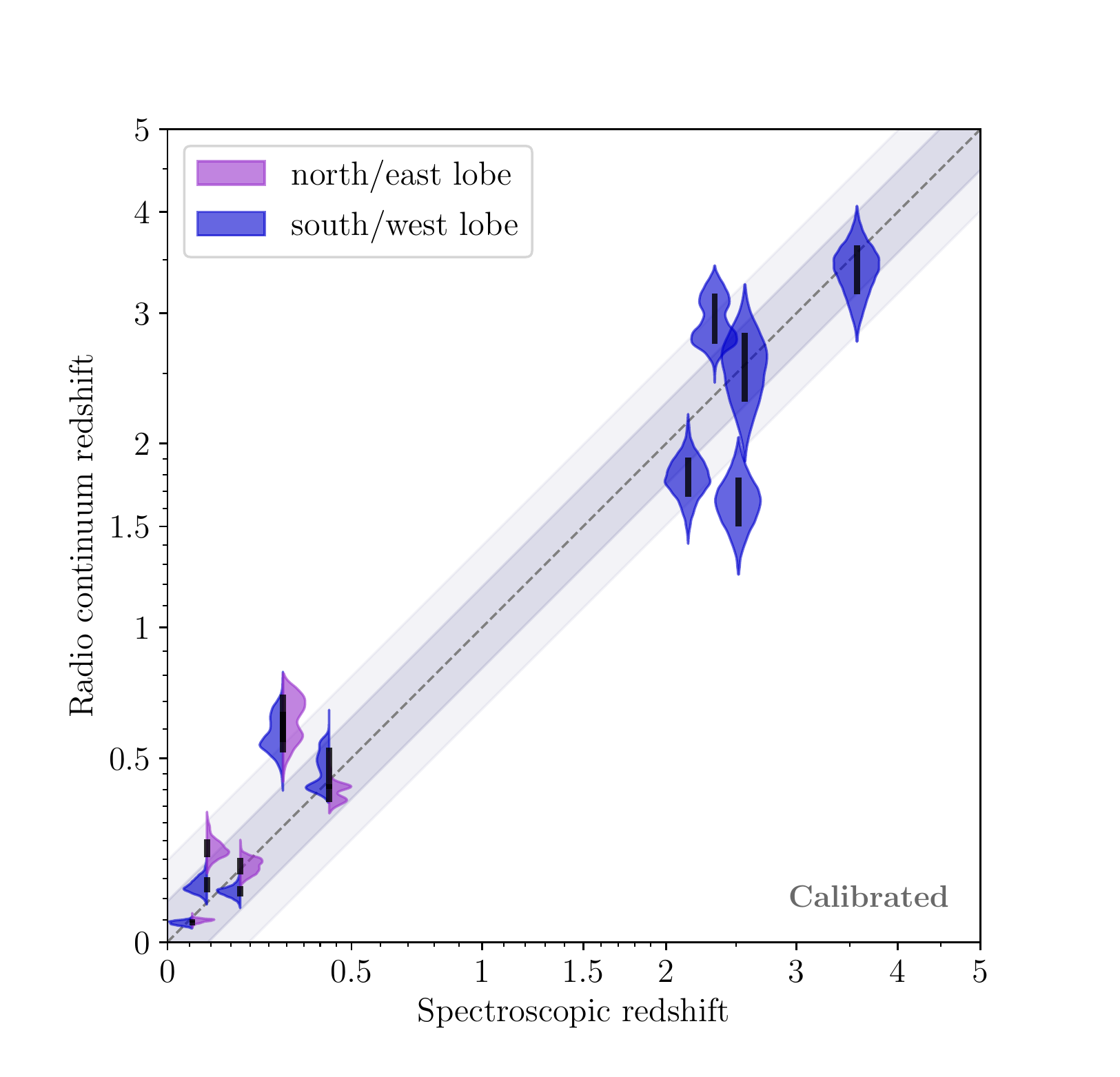}
\end{center}
\caption{Radio continuum redshifts for Cygnus A, 3C sources and high-redshift HeRGE objects as a function of their spectroscopic redshift. The five HeRGE radio AGNs have the same radio continuum redshift measurement for both lobes. The left figure shows the radio continuum redshifts for the uncalibrated model whilst the right plot is for the model calibrated based on the known spectroscopic redshifts. The shading indicates the spread of the redshift measurements about the one-to-one line (1 and 2$\sigma$ level shown). See the caption of Figure~\ref{fig:CygnusA} for a complete description of the plot.}
\label{fig:Herge_calib}
\end{figure*}

The \emph{RAiSE\textcolor{purple}{Red}} model is applied to the two lobes of Cygnus A, two lobes of the small subsample of 3C sources, and the combined lobes of each of the five HeRGE objects; i.e. 17 lobes are fitted with radio continuum redshifts. The redshift probability density functions for the majority of objects has a single sharp peak located approximately at the spectroscopic redshift. However, the probability density function for both lobes of 3C219 have two peaks, one at $z \approx 0.2$ and another at $z \approx 1.5$. The high-redshift peak is associated with lobe expansion rates at close to the speed of light and dissapears in the south lobe if the prior probability density function is modified to restrict velocities to be less than $0.5c$. That is, our technique can still find the correct radio continuum redshift for this source. The uncalibrated redshift distributions found using our technique for the remaining 15 lobes are plotted in Figure~\ref{fig:Herge_calib}(left) as a function of their known spectroscopic redshifts. The radio continuum redshifts are strongly correlated with the spectroscopic redshifts although the high-redshift HeRGE sample has their redshifts systematically underestimated. The uncalibrated model has an average log-space error of $\delta \log(1 + z^*) = 0.069\rm\, dex$ (i.e. 17\% in $1+z^*$). Importantly, the uncalibrated radio continuum redshifts explain 70\% of the variation in the known spectroscopic redshift for the high-redshift HeRGE sample, compared with at most 27\% for any one of the observables in isolation. 

The calibration constants are constrained (initally) using the radio continuum attributes of the 15 lobes (from Cygnus A, and the 3C and HeRGE subsamples). The fitted values are found as $b_1 = 0.76\pm0.07$, $b_2 = 0.01\pm0.01$, $b_3 = -0.60\pm0.07$ and $b_4 = -0.33\pm0.03$. These fits indicate that our parameter associated with the minimum Lorentz factor is a factor of 5.8 too low, the equipartition factor and gas density at redshift $z = 0$ are correct, the lobe to hotspot pressure ratio has a weaker than expected dependence on the axis ratio, and that the cosmological environments predict gas densities a factor of $(1 + z)^{0.33}$ too high. However, caution should be taken in directly associating these fits with their intended purpose (as stated above) as the least squares optimisation will also attempt to explain other more minor factors with these calibration constants. The redshift probability density functions for the 15 calibrator lobes are shown in Figure~\ref{fig:Herge_calib}(right) as a function of their spectroscopic redshifts. The correlation between these calibrated radio continuum redshifts and their spectrocopic redshifts is now centred on the one-to-one line with an average log-space error of $\delta \log(1 + z^*) = 0.040\rm\, dex$ (i.e. 9.6\% in $1+z^*$). Combining the probability density functions of the two lobes for each low-redshift source to yield a single robust radio continuum redshift does not change these results; however, the increased uncertainties on sources with less consistent estimates improves the agreement between these robust redshifts and their spectroscopic counterparts.

\begin{figure*}
\begin{center}
\includegraphics[width=0.42\textwidth,trim={25 20 55 45},clip]{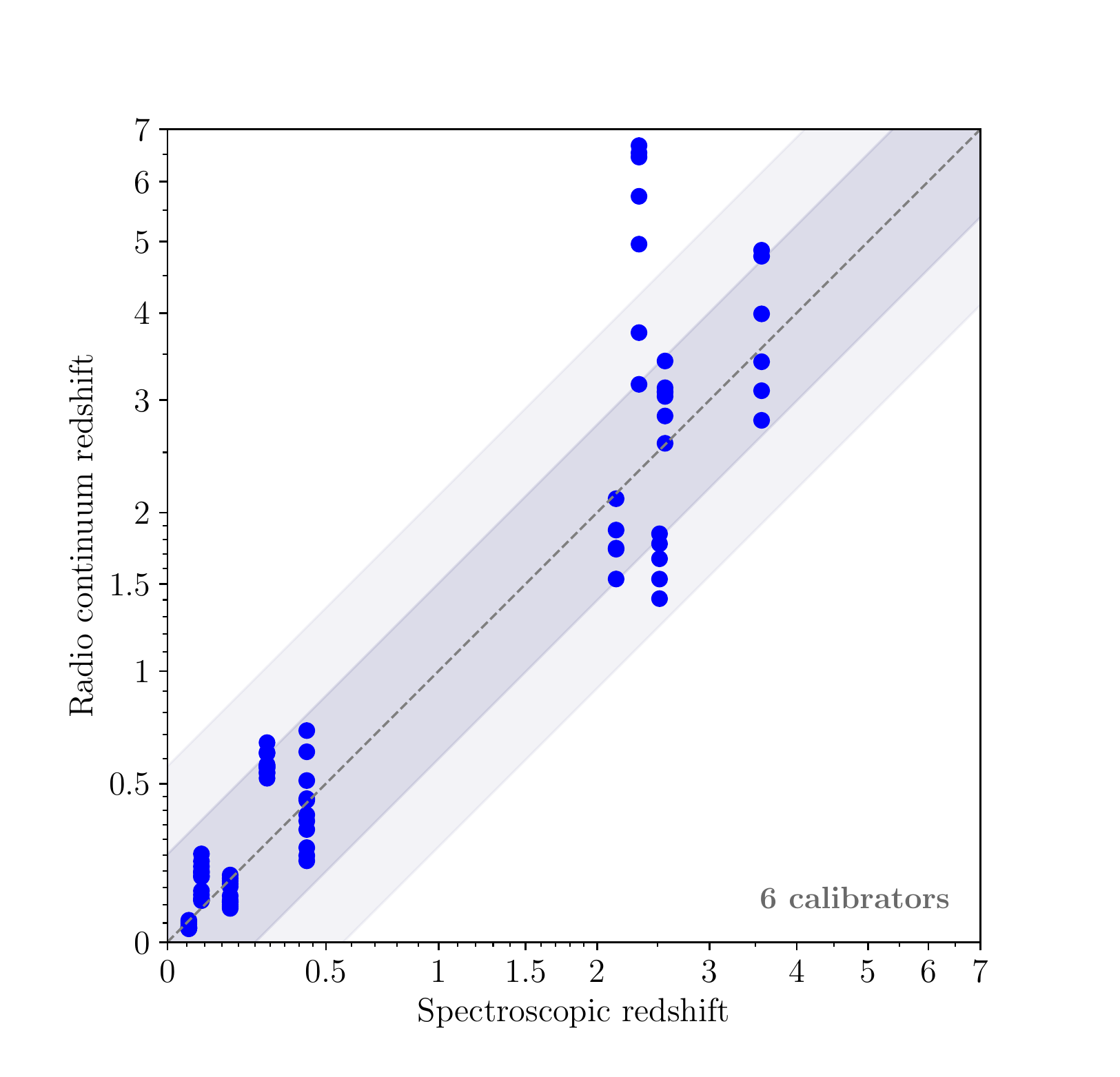}\quad\quad\quad\includegraphics[width=0.42\textwidth,trim={25 20 55 45},clip]{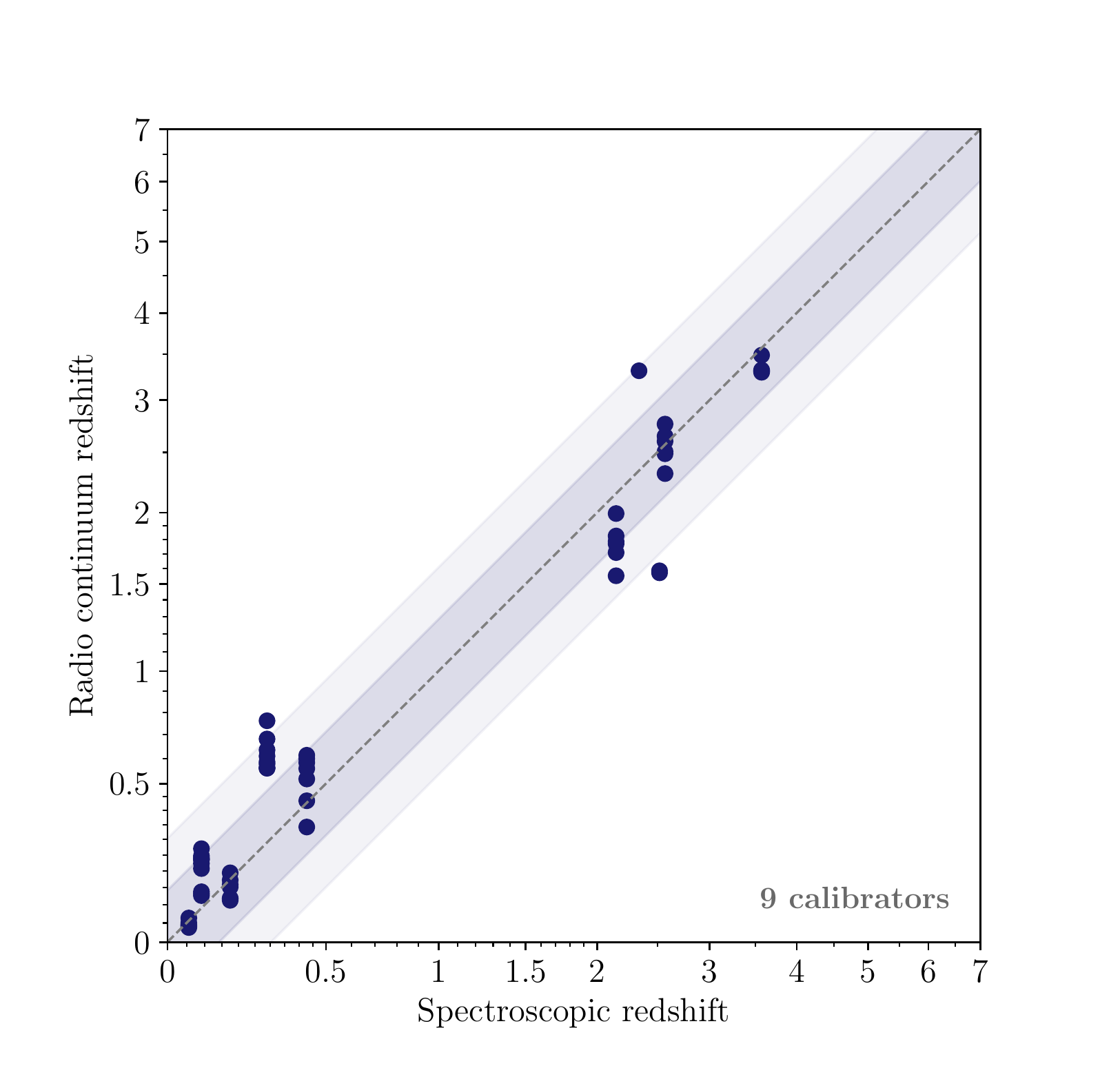}
\end{center}
\caption{Radio continuum redshifts for randomly selected subsamples of our low-redshift Cygnus A/3C sources and high-redshift HeRGE objects as a function of their spectroscopic redshift. The left figure shows radio continuum redshifts (mean values) for ten randomly selected subsamples (of nine objects) are calibrated using six calibrators whilst in the right plot the random subsamples (of six objects) are calibrated using nine calibrators. The shading indicates the spread of the redshift measurements about the one-to-one line (1 and 2$\sigma$ level shown). This graph is produced using a customised version of the \texttt{RAiSERed\_plotzz()} function.}
\label{fig:Herge_calib_sim}
\end{figure*}

We further quantify the likelihood that the calibration constants improve the accuracy of redshift measurements (compared to our uncalibrated model) for sources without known spectroscopic redshifts. This is achieved by randomly selecting six or nine objects as calibrators with the same ratio of low- and high-redshift lobes as our original sample (i.e. one-third HeRGE, two-thirds Cygnus A/3C); the remainder are specified to not have spectroscopic redshifts in the \emph{RAiSE\textcolor{purple}{Red}} code. Radio continuum redshifts are calculated for these remaining nine or six lobes respectively based on the calibration constants fitted using the randomly selected calibrators. This process is repeated ten times, selecting a different set of calibrators in each iteration. The redshifts estimated for the ten randomly selected subsets of lobes not used as calibrators are plotted in Figure~\ref{fig:Herge_calib_sim} as a function of their spectroscopic redshift. 

The radio continuum redshifts estimated when using six calibrators are largely consistent with their spectroscopic counterparts and centred on the one-to-one line. However, the predicted redshift for USS\,1707$+$105 is highly sensitive to the choice of calibrators, with estimates ranging from $z = 3$ to 7; the axis ratio upper limit is likely the cause of the increased variability in this source. As a result of this source, the likelihood of agreement with the spectrocopic redshifts actually decreases (compared to our uncalibrated model) when calibrating the model with six objects; the average error, across both redshift and the ten random source selections, increases to $\delta \log(1 + z^*) = 0.098\rm\, dex$ (i.e. 25\% in $1+z^*$)\footnote{The error is 14\% in $1+z^*$ if USS\,1707$+$105 is excluded from the calculation.}. 
By contrast, the radio continuum redshifts estimated using nine calibrators are much more stable due to the greater exploration of parameter space in the calibration. Variability in the estimated redshift of USS\,1707$+$105 is now minimised between calibrator samples. Calibration of the \emph{RAiSE\textcolor{purple}{Red}} model with nine sources increases the likelihood of agreement between the radio continuum and spectroscopic redshifts for all objects. The average error converges towards that found when using all 15 objects as calibrators; i.e. $\delta \log(1 + z^*) = 0.058\rm\, dex$ (i.e. 14\% in $1+z^*$). The error in our model is empirically related to the number of degrees of freedom, $\text{df} = N - 4$ (for $N$ calibrators), as $\delta \log(1 + z^*) = 0.139\;\!\text{df}^{\:\!-0.53}\rm\, dex$; i.e. the error is less than 10\% with 14 calibrators and, extrapolating, falls to less than 5\% with 40 sources. The proposed calibration technique will therefore be successful if a modest number of sources are used covering a comparable range of parameter space to the target sources lacking spectroscopic redshifts. 

\subsection{Ensemble verification metrics}

{There are a number of metrics used for optical photometric redshifts \citep[e.g.][]{Tanaka+2018, Schmidt+2020} that can similarly be applied to test the accuracy of the redshift probability density functions as an estimator of the true spectroscopic redshift. 
%In this work, we will consider a range of metrics based on the cumulative distribution function (CDF). 
In particular, the probability integral transformation (PIT) is the cumulative distribution function (CDF) of a redshift probability density function evaluated at its spectroscopic redshift:
\begin{equation}
\text{PIT} \equiv \text{CDF}(\tilde{p}, z^*) = \int_{-\infty}^{z^*} \tilde{p}(z) dz .
\end{equation}
The distribution of PIT values for accurate probability density functions is expected to be uniform from 0 to 1. An ensemble of overly broad probability density functions will produce an excess of PIT values around 0.5, and conversely, narrow distributions lead to an excess of the lowest and highest values. In this manner, the distribution of PIT values can be used to probe the average accuracy of the redshift probability density functions for an ensemble of radio AGNs.}

\begin{figure}
\begin{center}
\includegraphics[width=0.445\textwidth,trim={15 15 45 40},clip]{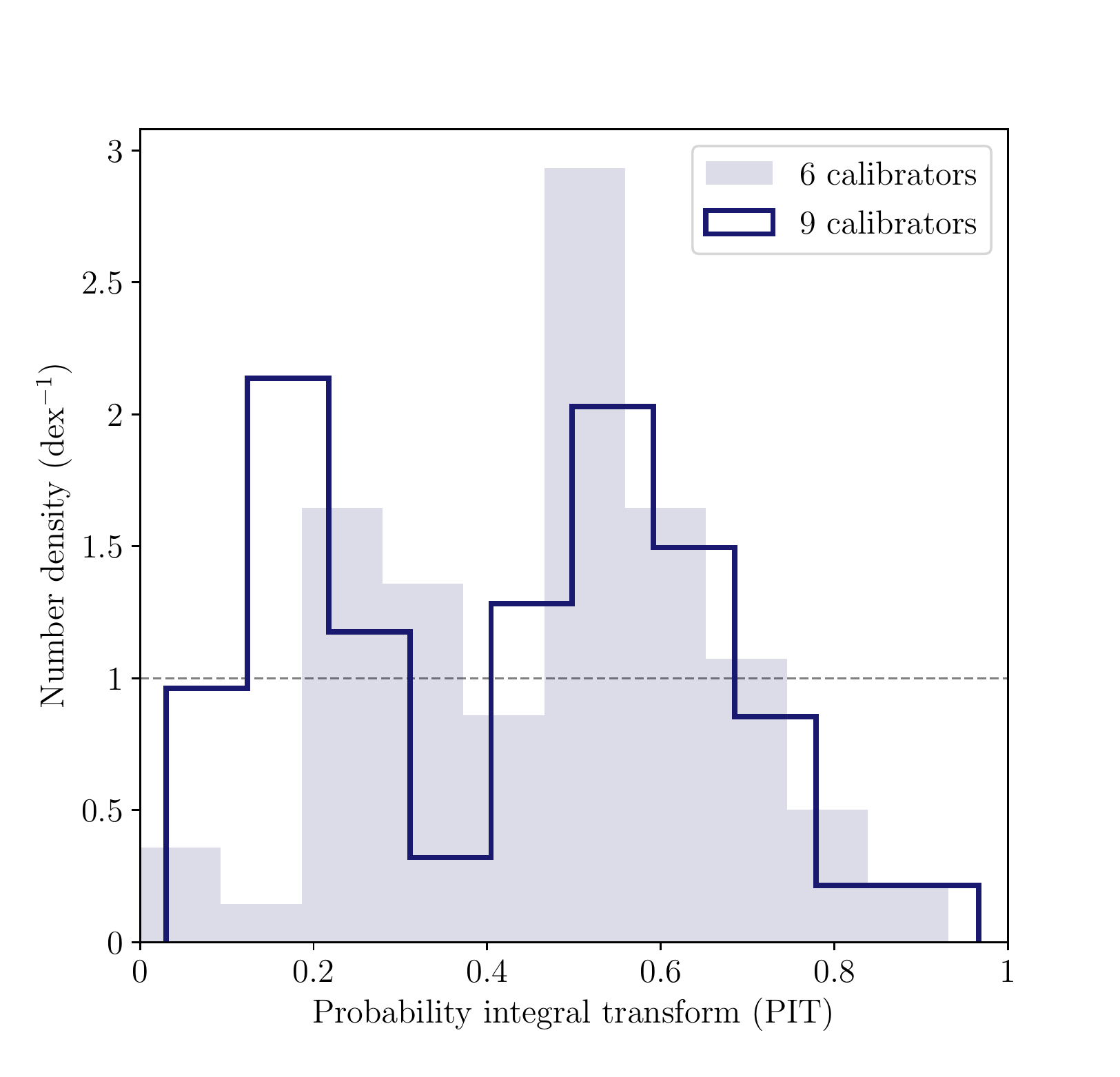}
\end{center}
\caption{{Histogram of the probability integral transform (PIT) of the redshift probability density functions for the random subsamples calibrated with either six (shaded) or nine (outlined) calibrators. The probability density functions used in this calculation include the systematic uncertainty of the model. The ideal PIT curve is shown by the horizontal dashed grey line.}}
\label{fig:stats_plot}
\end{figure}

{The accuracy of the redshift probability density functions for the calibrated subsamples considered in the previous section is assessed in Figure \ref{fig:stats_plot} using the probability integral transformation. These probability density functions are updated to include the model uncertainty of the \emph{RAiSE\textcolor{purple}{Red}} code. The distribution of the PIT values for the random subsamples (of six objects) calibrated using nine calibrators is evenly spread from 0 to 1, though perhaps is slightly right skewed.
Meanwhile, the distribution for the random subsamples (of nine objects) calibrated using six calibrators is drawn towards the centre (i.e. 0.5). Both sets of calibrated subsamples are reasonably consistent with the expected distribution of PIT values considering the size of the sample. Importantly, there are very few sources with extreme PIT values (e.g. $<$\,0.01 or $>$\,0.99) which would indicate outliers.}

{The distribution of PIT values for the two sets of subsamples are assessed quantitatively using the Kolmogorov-Smirnov and Anderson-Darling statistics. The Kolmogorov-Smirnov statistic measures the maximum difference between the CDF of the distribution of PIT values for the test subsample of radio AGNs and the CDF for the expected PIT distribution (i.e. uniform from 0 to 1). The Anderson-Darling statistic is a variant of the Kolmogorov-Smirnov statistic; see Equations 6 and 8 of \citet{Schmidt+2020}. The Kolmogorov-Smirnov statistic for the subsample using six calibrators is $K\!S=0.19$ ($p = 0.007$) and for the subsample with nine calibrators is $K\!S=0.22$ ($p = 0.016$); both have $p$-values of approximately 0.01 suggesting the PIT distributions are likely not consistent with the expected uniform distribution. Similarly, the Anderson-Darling statistics for the two sets of subsamples are $A\!D=8.3$ ($p = 0.001$) and $A\!D=4.1$ ($p = 0.007$) for six and nine calibrators respectively. By contrast, when considering the full sample of 15 calibrators the Anderson-Darling statistic drops to $A\!D=0.42$ ($p = 0.22$). The average accuracy of the \emph{RAiSE\textcolor{purple}{Red}} redshift estimates therefore becomes consistent with the expected PIT distribution if a sufficient number of objects are used in the calibration.}

\section{CONCLUSIONS}
\label{sec:CONCLUSIONS}

We have presented an algorithm, based on the active galactic nucleus (AGN) standard candles of \citet{Turner+2019}, that generates redshift probability density functions using only radio-frequency imaging and photometry of extended radio AGNs.
Specifically, our \emph{RAiSE\textcolor{purple}{Red}} model uses five attributes measured from the radio-frequency observations to find the most likely redshift assuming a prior probability density function for the density profile of the cosmological environments; fixed values are assumed for other model parameters including the minimum Lorentz factor of the electron energy distribution and the ratio of the energy density in the magnetic field to that in the particles (i.e. equipartition factor). The observed attributes are measured individually for each lobe; they are: (i) the angular size, $\theta$; (ii) angular width, $w = 2\theta/A$ for axis ratio $A$; (iii) the `optically thin' spectral break frequency, $\nu_{\rm b}$; (iv) the spectral index below the break frequency, $\alpha_{\rm inj} = (s - 1)/2$; and (v) the integrated flux density, $S_\nu$, at some frequency below the break (i.e. $\nu \ll \nu_{\rm b}$). 

We create a mock radio source population based on the properties of Cygnus A to assess the sensitivity of our \emph{RAiSE\textcolor{purple}{Red}} algorithm to large uncertainties or upper/lower bounds on the five observed attributes. We find that upper bounds on the angular size, as in unresolved sources, are sufficient to yield accurate radio continuum redshift measurements at $z \geqslant 2$; at lower redshifts the angular size is a critical constraint. Meanwhile, the break frequency can have moderate uncertainties of $0.5\rm\, dex$ without affecting the estimated radio continuum redshifts. Radio sources with break frequencies below the range of observing frequencies (i.e. aged-spectrum with $\alpha > 1$) can have accurate redshifts measured up to $z \leqslant 1$, whilst those with break frequencies above the observed frequencies (i.e. $0.5 < \alpha < 1$) have accurate measurements for redshifts above $z \geqslant 3$.

The \emph{RAiSE\textcolor{purple}{Red}} model is applied to a sample of 17 radio AGN lobes comprising Cygnus A, objects from the low-redshift ($z < 1$) \emph{Third Cambridge Catalogue of Radio Sources} (3C), and a subsample of the high-redshift ($2 < z < 4$) \emph{Herschel Radio Galaxy Evolution} project (HeRGE). All but two of the lobes have a single, well-defined peak in their probability density functions; 3C219 has peaks at two redshifts, one corresponding to the known spectroscopic redshift of $z = 0.1744$ and another at $z \approx 1.5$; the second peak is ruled out by imposing a maximum lobe expansion velocity of $0.5c$. The radio continuum redshifts derived for our uncalibrated model have an average error of 17\% (in redshift as $1 + z^*$) compared to their spectroscopic measurement. Importantly, the redshifts derived by our uncalibrated algorithm explain 70\% of the variation in the spectroscopic redshifts of the high-redshift HeRGE sample, compared to at most 27\% for any one of the observed attributes in isolation. That is, the model performs significantly better than a simple redshift--flux density or redshift--angular size relationship by considering the intrinsic physics of the radio AGNs.

We calibrate the most poorly constrained parameters in our model using a modest sample (6-9 sources) with known spectroscopic redshifts. These properties include the minimum Lorentz factor, gas density at the working surface of the lobe, equipartition factor, lobe to hotspot pressure ratio as a function of axis ratio, and the evolution of cosminc environments with redshift. The uncertainty in these parameters are captured in four calibration constants, $b_1$, $b_2$, $b_3$ and $b_4$. The error in the \emph{RAiSE\textcolor{purple}{Red}} model upon calibration using nine sources with known spectroscopic redshifts reduces to 14\% (in redshift as $1 + z^*$) across all redshifts. %The error further reduces to 9.6\% if, slightly unrealistically, all the objects in our sample are used to calibrate the model. 
The calibration of our algorithm is therefore expected to yield improved (and accurate) results compared to our uncalibrated model using our best guesses for the poorly constrained parameters.

Next-generation radio surveys are expected to observe tens of millions of AGNs with a median redshift in excess of $z \geqslant 1$. For example, the \emph{Evolutionary Map of the Universe} (EMU) aims to detect about 70 million sources, about half of which are expected to be star-forming galaxies and the rest AGNs \citep{Norris+2019}. Radio-frequency imaging and broadband photometry covering much of the frequency range from 10-1800\,MHz will be available in both the northern and southern skies with ASKAP EMU \citep{Norris+2011}, ASKAP POSSUM \citep{Gaensler+2010}, LOFAR LoTSS \citep{Shimwell+2017, Shimwell+2019}, MeerKAT MIGHTEE \citep{Jarvis+2012}, MWA GLEAM \citep{Wayth+2015}, and VLA VLASS \citep{Lacy+2020}. There is therefore sufficient data for our \emph{RAiSE\textcolor{purple}{Red}} model to be applied to any presently-active extended AGNs with a lobed morphology identified in these surveys. Crucially, beyond targeted surveys, the vast majority of these objects will not have spectroscopic redshifts, whilst photometric redshifts for high-redshift AGNs are expected to be of limited quality, and even then require optical and infrared photometry. Radio continuum redshifts are therefore likely to be a valuable tool in investigating the properties of the very numerous AGN population, whilst also offering a means to provide indirect redshift estimates to other galaxies identified as belonging to the same projected structures by machine learning algorithms.  We provide \emph{python} code for the calculation, calibration and plotting of our radio continuum redshifts and their probability density functions in the online supplementary material.

\subsection*{Data availablity}

The authors confirm that the data supporting the findings of this study are available within the article and the relevant code is included in the supplementary materials.
\newline

\noindent
We thank an anonymous referee for helpful and constructive comments that have improved our manuscript.

%\begin{appendix}
%\onecolumn
%\section{Code documentation and examples}
%\label{sec:appendix}

%\begin{lstlisting}
%import RAiSERed as rr
%array = rr.RAiSERed_list()
%array.append(rr.RAiSERed('Cygnus A', [0.151], [5960], 58.57, 9.25, 2.5, 2.78, specz=0.056075))
%\end{lstlisting}

%\end{appendix}


\begin{thebibliography}{}

\bibitem[Alexander(2000)]{Alexander+2000}
Alexander, P. 2000, MNRAS, 319, 8

\bibitem[Amaro(2018)]{Amaro+2018}
Amaro, V., Cavuoti, S., Brescia, M., et al. 2018, MNRAS, 482, 3116

\bibitem[Arnouts et al.(1999)]{Arnouts+1999}
Arnouts, S., Cristiani, S., Moscardini, L., Matarrese, S., Lucchin, F., et al., 1999, MNRAS, 310, 540

\bibitem[Blundell et al.(1999)]{Blundell+1999}
Blundell, K. M., Rawlings, S., \& Willott, C. J. 1999, AJ, 117, 677

\bibitem[Broderick et al.(2007)]{Broderick+2007}
Broderick, J. W., De Breuck, C., Hunstead, R. W., \& Seymour, N. 2007, MNRAS, 375, 1059

\bibitem[Buchalter et al.(1998)]{Buchalter+1998}
Buchalter, A., Helfand, D. J., Becker, R. H., \& White, R. L. 1998, ApJ, 494, 503

\bibitem[Carilli et al.(1996)]{Carilli+1996}
Carilli, C. L., \& Barthel, P. D. 1996, ApJ, 383, 554

\bibitem[Carilli et al.(1991)]{Carilli+1991}
Carilli, C. L., Perley, R. A., Dreher, J. W., \& Leahy, J. P., 1991, ApJ, 383, 554

\bibitem[Carilli et al.(1997)]{Carilli+1997}
Carilli, C. L., R\"{o}ttgering, H. J. A., van Ojik, R., Miley, G. K., \& van Breugel, W. J. M. 1997, ApJS, 109, 1

\bibitem[Cohen et al.(2007)]{Cohen+2007}
Cohen, A. S., Lane, W. M., Cotton, W. D., et al. 2007, AJ, 134, 1245

\bibitem[Croton et al.(2006)]{Croton+2006}
Croton, D. J., Springel, V., White, S. D. M., et al. 2006, MNRAS, 365, 11

\bibitem[Croton et al.(2016)]{Croton+2016}
Croton, D. J., Stevens, A. R. H., Tonini, C., et al. 2016, ApJS, 222, 22

\bibitem[Czerny et al.(2013)]{Czerny+2013}
Czerny, B., Hryniewicz, K., Maity, I., et al. 2013, A\&A, 556, 97

\bibitem[Daly(1994)]{Daly+1994}
Daly, Ruth A. 1994, ApJ, 426, 38

\bibitem[De Breuck et al.(2010)]{DeBreuck+2010}
De Breuck, C., Seymour, N., Stern, D., et al. 2010, ApJ, 725, 36

\bibitem[Drouart et al.(2014)]{Drouart+2014}
Drouart, G., De Breuck, C., Vernet, J., et al. 2014, A\&A, 566, 53

\bibitem[Drouart et al.,(in preparation)]{Drouart+2020}
Drouart, G., et al., in preparation

\bibitem[Duncan et al.(2018a)]{Duncan+2018a}
Duncan, K. J., Brown, M. J. I., Williams, W. L., et al. 2018a, MNRAS, 473, 2655

\bibitem[Duncan et al.(2018b)]{Duncan+2018b}
Duncan, K. J., Jarvis, M. J., Brown, M. J. I., \& R\"{o}ttgering, H. J. A. 2018b, MNRAS, 477, 5177

\bibitem[Fabian(2012)]{Fabian+2012}
Fabian, A. C. 2012, ARA\&A, 50, 455

\bibitem[Fanaroff \& Riley(1974)]{FR+1974}
Fanaroff, B. L., \& Riley, J. M. 1974, MNRAS 167, 31

\bibitem[Firth et al.(2002)]{Firth+2002}
Firth, A.E., Lahav, O. and Somerville, R. S., 2002, MNRAS

\bibitem[Franzen et al.,(in preparation)]{Franzen+2020}
Franzen, T. M. O., et al., in preparation

\bibitem[Freedman et al.(2019)]{Freedman+2019}
Freedman, W. L., Madore, B, F., Hatt, D., et al. 2011, ApJ, 882, 34

\bibitem[Gaensler et al.(2010)]{Gaensler+2010}
Gaensler, B. M., Landecker, T. L., Taylor, A. R., \& POSSUM Collaboration 2010, Bulletin of the American Astronomical Society, 42, 470.13

\bibitem[Girardi \& Giuricin(2000)]{Girardi+2000}
Girardi, M., \& Giuricin, G. 2000, ApJ, 540, 45

\bibitem[Gonzalez et al.(2013)]{Gonzalez+2013}
Gonzalez, A. H., Sivanandam, S., Zabludoff, A. I., \& Zaritsky, D. 2013, ApJ, 778, 14

\bibitem[Haas et al.(2011)]{Haas+2011}
Haas, M., Chini, R., Ramolla, M., et al. 2011, A\&A, 535, 73

\bibitem[Hardcastle(2018)]{Hardcastle+2018}
Hardcastle, M. J. 2018, MNRAS, 475, 2768

\bibitem[Hardcastle \& Krause(2014)]{Hardcastle+2014}
Hardcastle, M. J., \& Krause, M. G. H. 2014, MNRAS, 443, 1482

\bibitem[Harwood(2017)]{Harwood+2017}
Harwood, J. J. 2017, MNRAS, 466, 2888

\bibitem[H\"{o}nig et al.(2017)]{Honig+2017}
H\"{o}nig, S. F., Watson, D., Kishimoto, M., et al. 2017, MNRAS, 464, 1693

\bibitem[H\"{o}nig et al.(2014)]{Honig+2014}
H\"{o}nig, S. F., Watson, D., Kishimoto, M., \& Hjorth, J. 2014, Nature, 515, 528

\bibitem[Hurley-Walker et al.(2017)]{Hurley-Walker+2017}
Hurley-Walker, N.,  Callingham, J. R., Hancock, P. J., et al. 2017, MNRAS, 464, 1146

\bibitem[Jackson(2004)]{Jackson+2004}
Jackson, J. C. 2004, JCAP, 11, 7

\bibitem[Jarvis(2012)]{Jarvis+2012}
Jarvis, M. J. 2012, AfrSk, 16, 44

\bibitem[Kaiser(2000)]{Kaiser+2000}
Kaiser, C. R. 2000, A\&A, 362, 447

\bibitem[Kaiser \& Alexander(1997)]{KA+1997}
Kaiser, C. R., \& Alexander, P. 1997, MNRAS, 286, 215

\bibitem[Kaiser \& Alexander(1999)]{Kaiser+1999}
Kaiser, C. R., \& Alexander, P. 1999, MNRAS, 305, 707

\bibitem[Kellermann(1993)]{Kellermann+1993}
Kellermann, K. I. 1993, Nature, 361, 134

\bibitem[King et al.(2014)]{King+2014}
King, A. L., Davis, T. M., Denney, K. D., Vestergaard, M., \& Watson, D. 2014, MNRAS, 441, 3454

\bibitem[Klypin et al.(2011)]{Klypin+2011}
Klypin, A., Trujillo-Gomez, S., \& Primack, J. 2011, ApJ, 740, 102

\bibitem[Krause et al.(2012)]{Krause+2012}
Krause, M., Alexander, P., Riley, J., \& Hopton, D. 2012, MNRAS, 427, 3196

\bibitem[Lacy et al.(2020)]{Lacy+2020}
Lacy, M., Baum, S. A., Chandler, C. J., et al. 2020, PASP, 132, 1009

\bibitem[Laing \& Peacock(1980)]{Laing+1980}
Laing, R. A., \& Peacock, J. A. 1980, MNRAS, 190, 903

\bibitem[Laing et al.(1983)]{Laing+1983}
Laing, R. A., Riley, J. M., \& Longair, M. S. 1983, MNRAS, 204, 151

\bibitem[Massaglia et al.(2019)]{Massaglia+2019}
Massaglia, S.,  Bodo, G., Rossi, P., Capetti, S., \& Mignone, A. 2019, A\&A, 621, 132

\bibitem[Mayer et al.(2016)]{Mayer+2016}
Mayer, A.,  Mora, T., Rivoire, O., \& Walczak, A. M. 2016, PNAS, 113, 8630

\bibitem[McGaugh et al.(2010)]{McGaugh+2010}
McGaugh, S. S., Schombert, J. M., de Blok, W. J. G., \& Zagursky, M. J. 2010, ApJ, 708, L14

\bibitem[Mullin et al.(2008)]{Mullin+2008}
Mullin, L. M., Riley, J. M., \& Hardcastle, M. J. 2008, MNRAS, 390, 595

\bibitem[Norris(2017)]{Norris+2017}
Norris, R. P. 2017, Nature Astronomy, 1, 671

\bibitem[Norris et al.(2011)]{Norris+2011}
Norris, R. P., et al. 2011, PASA, 28, 215

\bibitem[Norris et al.(2013)]{Norris+2013}
Norris, R. P., et al. 2013, PASA, 30, e020

\bibitem[Norris et al.(2019)]{Norris+2019}
Norris, R. P., Salvato, M., Longo, G., et al. 2019, PASP, 131, 1004

\bibitem[Oknyanskij et al.(1999)]{Oknyanskij+1999}
Oknyanskij, V. L., Lyuty, V. M., Taranova, O. G., \& Shenavrin, V. I. 1999, AstL, 25, 483

\bibitem[Owen et al.(1997)]{Owen+1997}
Owen, F. N., Ledlow, M. J., Morrison, G. E., \& Hill, J. M. 1997, ApJ, 488, 15

\bibitem[Pentericci et al.(2001)]{Pentericci+2001}
Pentericci, L., McCarthy, P. J., R\"{o}ttgering, H. J. A., Miley, G. K., van Breugel, W. J. M., \& Fosbury, R. 2001, ApJS, 135, 63

\bibitem[Pentericci et al.(2000)]{Pentericci+2000}
Pentericci, L., Van Reeven, W., Carilli, C. L., R\"{o}ttgering, H. J. A., \& Miley, G. K. 2000, A\&AS, 145, 121

\bibitem[Planck Collaboration(2016)]{Planck+2016}
Planck Collaboration 2016, A\&A, 594, A13

\bibitem[Pope et al.(2012)]{Pope+2012}
Pope, E. C. D., Mendel, T., \& Shabala, S. S. 2012, MNRAS, 419, 50

\bibitem[Raouf et al.(2017)]{Raouf+2017}
Raouf, M., Shabala. S.~S., Croton, D.~J., Khosroshahi, H.~G., \& Bernyk, B. 2017, MNRAS, 471, 658

\bibitem[Riess et al.(2016)]{Riess+2016}
Riess, A. G., Marci, L. M., Hoffmann, S. L., et al 2016, ApJ, 826, 56

\bibitem[Riess et al.(2019)]{Riess+2019}
Riess, A. G., Casertano, S., Yuan, W., Marci, L. M., \& Scolnic, D. 2019, ApJ, 876, 85

\bibitem[Sabater et al.(2019)]{Sabater+2019}
Sabater, J., Best. P.~N., Hardcastle, M.~J., et al. 2019, A\&A, 622A, 17

\bibitem[Salvato et al.(2018)]{Salvato+2018}
Salvato, M., Ilbert, O., \& Hoyle, B., 2018, Nature Astronomy, 3, 212

\bibitem[Seymour et al.(2007)]{Seymour+2007}
Seymour, N., Stern, D., De Breuck, C., et al. 2007, ApJS, 171, 353

\bibitem[Seymour et al.,(in preparation)]{Seymour+2020}
Seymour, N., et al., in preparation

\bibitem[Schmidt et al.(2020)]{Schmidt+2020}
Schmidt, S. J., Malz, A. I., Soo, J. Y. H., et al. 2020, MNRAS, arXiv:2001.03621

\bibitem[Shimwell et al.(2017)]{Shimwell+2017}
Shimwell, T. W., R\:{o}ttgering, H. J. A., Best, P. N., et al. 2017, A\&A, 598, 104

\bibitem[Shimwell et al.(2019)]{Shimwell+2019}
Shimwell, T. W., Tasse, C., Hardcastle, M. J., et al. 2019, A\&A, 622

\bibitem[Steenbrugge et al.(2010)]{Steenbrugge+2010}
Steenbrugge, K. C., Keywood, I. \& Blundell, K. M. 2010, MNRAS, 401, 67

\bibitem[Tagliaferri et al.(2003)]{Tagliaferri+2003}
Tagliaferri, R., Longo, G., Andreon, S., et al. 2003, Lecture Notes in Computer Science, 2859, 226

\bibitem[Tanaka et al.(2018)]{Tanaka+2018}
Tanaka, M., Coupon, J., Hsieh, B-C., et al. 2018, PASJ, 70, S9

\bibitem[Turner(2018)]{Turner+2018}
Turner, R. J. 2018, MNRAS, 476, 2522

\bibitem[Turner et al.(2018a)]{Turner+2018a}
Turner, R. J., Rogers, J. G., Shabala, S. S., \& Krause, M. G. H. 2018a, MNRAS, 473, 4179

\bibitem[Turner \& Shabala(2015)]{Turner+2015}
Turner, R. J., \& Shabala, S. S. 2015, ApJ, 806, 59

\bibitem[Turner \& Shabala(2019)]{Turner+2019}
Turner, R. J., \& Shabala, S. S. 2019, MNRAS, 486, 1225

\bibitem[Turner \& Shabala(2020)]{Turner+2020}
Turner, R. J., \& Shabala, S. S. 2020, MNRAS, 493, 5181

\bibitem[Turner et al.(2018b)]{Turner+2018b}
Turner, R. J., Shabala, S. S., \& Krause, M. G. H. 2018b, MNRAS, 474, 3361

\bibitem[van Ojik et al.(1996)]{van Ojik+1996}
van Ojik, R., R\"{o}ttgering, H. J. A., Carilli, C. L., Miley, G. K., Bremer, M. N., \& Macchetto, F. 1996, A\&A, 313, 25

\bibitem[Vikhlinin et al.(2006)]{Vikhlinin+2006}
Vikhlinin, A., Kravtsov, A., Forman, W., et al. 2006, ApJ, 640, 691

\bibitem[Watson et al.(2011)]{Watson+2011}
Watson, D., Denney, K. D., Vestergaard, M., \& Davis, T. M. 2011, ApJ, 740, L49

\bibitem[Wayth et al.(2015)]{Wayth+2015}
Wayth, R. B., Lenc, E., Bell, M. E., et al. 2015, PASA, 32, 25

\bibitem[Yates et al.(2018)]{Yates+2018}
Yates, P. M., Shabala, S. S., \& Krause, M. G. H. 2018, MNRAS, 480, 5286

\bibitem[Yoshii et al.(2014)]{Yoshii+2014}
Yoshii, Y., Kobayashi, Y., Minezaki, T., Koshida, S., \& Peterson, B. A. 2014, ApJ, 784, L11

\end{thebibliography}
\end{document}